\documentclass[fleqn,usenatbib]{mnras}
\usepackage{newtxtext,newtxmath}
\usepackage[flushleft]{threeparttable}

\usepackage[T1]{fontenc}
\usepackage{ae,aecompl} 

\usepackage{graphicx}	
\usepackage{caption}

\title[A Comparison of CGM Around Dwarf Galaxies]{Probing the CGM of Low-redshift Dwarf Galaxies Using FIRE Simulations}

\author[F. Li et al.]{Fei Li,$^{1,2}$\thanks{E-mail: lif.li@mail.utoronto.ca (FL)}
Mubdi Rahman,$^{1,3}$
Norman Murray,$^{2,4}$, 
Zachary Hafen,$^{5}$
 \newauthor Claude-Andr{\'e} Faucher-Gigu{\`e}re$^{5}$, Jonathan Stern$^{5}$, Cameron B. Hummels$^{6}$, \newauthor Philip F. Hopkins$^{6}$, Kareem El-Badry$^{7}$, and Du{\v s}an Kere{\v s}$^{8}$\\
$^{1}$David A. Dunlap Department of Astronomy and Astrophysics, University of Toronto, 50 St. George Street, ON M5S 3H4, Canada.\\
$^{2}$Canadian Institute for Theoretical Astrophysics, University of Toronto, 60 St. George Street, Toronto, ON M5S 3H8, Canada.\\
$^{3}$Dunlap Institute for Astronomy \& Astrophysics, University of Toronto, 50 St. George St., Toronto, ON M5S 3H4, Canada.\\
$^{4}$Canada Research Chair in Astrophysics.\\
$^{5}$Department of Physics and Astronomy and CIERA, Northwestern University, 2145 Sheridan Road, Evanston, IL 60208, USA.\\
$^{6}$TAPIR, Mailcode 350-17, California Institute of Technology, Pasadena, CA 91125, USA.\\
$^{7}$Department of Astronomy and Theoretical Astrophysics Center, University of California Berkeley, Berkeley, CA 94720, USA.\\
$^{8}$Department of Physics, Center for Astrophysics and Space Sciences, University of California at San Diego, 9500 Gilman Drive, \\ 
La Jolla, CA 92093, USA. \\
}

\date{Accepted XXX. Received YYY; in original form ZZZ}

\pubyear{2020}

\begin{document}

\label{firstpage}
\pagerange{\pageref{firstpage}--\pageref{lastpage}}
\maketitle

\begin{abstract}
Observations of UV metal absorption lines have provided insight into the structure and composition of the circumgalactic medium (CGM) around galaxies. We compare these observations with the low-redshift ($z \leq 0.3$) CGM around dwarf galaxies in high-resolution cosmological zoom-in runs in the FIRE-2 simulation suite. We select simulated galaxies that match the halo mass, stellar mass, and redshift of the observed samples. We produce absorption measurements using \textsc{Trident} for UV transitions of \ion{C}{IV}, \ion{O}{VI}, \ion{Mg}{II} and \ion{Si}{III}. The FIRE equivalent width (EW) distributions and covering fractions for the \ion{C}{IV} ion are broadly consistent with observations inside $0.5 R_{\rm vir}$, but are under-predicted for \ion{O}{VI}, \ion{Mg}{II}, and \ion{Si}{III}. The absorption strengths of the ions in the CGM are moderately correlated with the masses and star formation activity of the galaxies. The correlation strengths increase with the ionization potential of the ions. The structure and composition of the gas from the simulations exhibit three zones around dwarf galaxies characterized by distinct ion column densities: the disky ISM, the inner CGM (the wind-dominated regime), and the outer CGM (the IGM accretion-dominated regime). We find that the outer CGM in the simulations is nearly but not quite supported by thermal pressure, so it is not in hydrostatic equilibrium (HSE), resulting in halo-scale bulk inflow and outflow motions. The net gas inflow rates are comparable to the SFR of the galaxy, but the bulk inflow and outflow rates are greater by an order of magnitude, with velocities comparable to the virial velocity of the halo. These roughly virial velocities (${\sim} 100 \,\rm km\,s^{-1}$) produce large EWs in the simulations. This supports a picture for dwarf galaxies in which the dynamics of the CGM at large scales are coupled to the small-scale star formation activity near the centre of their halos. 
\end{abstract}



\section{Introduction}



The origin of the circumgalactic medium (CGM) is a product of two critical processes in galaxy evolution: gas inflow from the intergalactic medium (IGM) and gas outflow from the galaxy. Continuous gas inflow is required to sustain the star formation activity observed throughout cosmic time, while galactic outflows are necessary to explain the low observed star formation efficiency \citep{Dekel, Silk}. However, the CGM is a complex and dynamic environment because of the interaction between accretion and outflows, and hydrodynamic instabilities that occur in the process. Consequently, the study of the CGM is both challenging and crucial for our understanding of galaxy formation.

The CGM is diffuse and dim, and as a result, it has only recently been detected in emission \citep[e.g.,][]{Martin2014, Martin2015}. An alternate probe is to use absorption spectra of bright background objects (e.g., quasars) to gain information on element abundances along sightlines. With recent observations probing the circumgalactic region \citep{Tumlinson2011, Prochaska2011, Werk2014, Bordoloi2014, Liang2014, Johnson2015, Johnson2017}, we are provided with new constraints to test and inform our theoretical and computational modeling of the process of galaxy formation and evolution. By comparing the structure of the CGM in simulations with that seen in observations, we can test our understanding of galaxy formation. At the same time, because of the limitations of the observational probes, the simulations can provide an interpretation of the observations. 

Small scale simulations with sufficient resolution to resolve individual stars have identified non-linear interaction of the different feedback mechanisms as critical for the generation of outflows from stars. However, such extremely high resolution simulations are restricted to small scale systems such as single star clusters, molecular clouds \citep[e.g.,][]{Harper-Clark2011} or the ``first stars" \citep[e.g.,][]{Wise2012}. Current large-scale cosmological simulations \citep[e.g.,][]{Vogelsberger2014a, Vogelsberger2014}, still cannot resolve the details of the launching of stellar winds from stellar feedback and thus need to implement ``sub-grid" prescriptions. The prescriptions are routinely tuned so that the simulations match the observational constraints, such as the stellar mass -- halo mass function. 

The FIRE (Feedback In Realistic Environments) project \citep{Hopkins2014, Hopkins2018} includes a suite of high-resolution cosmological zoom-in simulations \footnote{\url{http://fire.northwestern.edu/}}. What distinguishes the simulations in the FIRE project from other cosmological simulations is the realistic feedback models used. The feedback mechanisms included in the simulation suite in this analysis are supernovae (Ia \& II), continuous stellar mass-loss (OB/AGB-star winds), photo-ionization and photo-electric heating, radiation pressure, metal diffusion, and cosmic rays. The parameters in the feedback models are not tuned ``by hand" to match the observational constraints. Instead, the stellar feedback inputs are based entirely on stellar population models. It has already been shown in a series of papers that the FIRE simulations can reproduce a number of galaxy properties in observations at different redshifts, including star-formation \citep[e.g.,][]{Feldmann2016, Sparre2017, Orr2018}, dynamical \citep[e.g.,][]{Orr2019} and morphological properties \citep[e.g.,][]{Ma, Wetzel, El-Badry2018}. Results discussed later in this paper include comparison of FIRE-2 simulations with and without cosmic rays to observations of the CGM around low-redshift ${\sim} L \star$ galaxies \citep{Ji}, particle tracking of inflows and outflows that build up galaxies \citep{Angles2017}, and particle tracking of origins and fates of the CGM \citep{Hafen2019b, Hafen2019}.

The FIRE feedback implementation makes clear predictions of the dynamics and thermal state of the CGM in the main halo of each simulation. 
The predictions of neutral hydrogen absorption in the CGM of the FIRE simulations have been presented in \citet{Faucher-Giguere2015, Faucher-Giguere2016} and \citet{Hafen2017}, so here we concentrate on metal absorption lines.

In this work, our goal is to compare the structure of the CGM around low-redshift dwarf galaxies seen in the FIRE-2 simulations with the results of the available observations \citep{Bordoloi2014,Liang2014, Johnson2017}. Following \citet{Bordoloi2014}, who include galaxies with stellar masses up to $1.6 \times 10^{10} M_{\rm \odot}$, we include galaxies up to $2.5 \times 10^{10} M_{\rm \odot}$ (FIRE simulation `m11f') to cover the observed range.

With shallower potential wells and higher gas fractions, dwarf galaxies are both predicted and observed \citep{Martin1998, Martin2002, Strickland2000, Summers2003, Summers2004, Hartwell2004, Ott2005} to have stronger outflows relative to their stellar mass or star formation rates. Because of the difference in the order of magnitude in mass, dwarf galaxies provide a critical testing ground for our theories of galaxy evolution and a great opportunity to understand the process of the cycling of baryons through galactic disks into the CGM and back over a broad range of halo masses. We leave comparison of high-z dwarf galaxies and  predictions of the distributions of other ions for future work. 

This paper is structured as follows. In Section \ref{sec:simulations}, we describe the simulation suite and the observations used in this study. In Section \ref{sec:post-processing}, we describe the post-processing of the simulations for CGM analysis. In Section \ref{sec:results}, we present the main results from the analysis of the simulation data, including the absorption EW distributions of four ions in the simulated sample and their comparison with observations (Section \ref{sec:EW}), examination of the relationship between the equivalent width (EW) distributions of the ions in the simulations (Section \ref{sec:CGM_phase_space}), connecting UV absorption to galactic properties (Section \ref{sec:global_properties}), physical distribution of the CGM gas (Section \ref{sec:physical_dist}), and the dynamics of the CGM around dwarf galaxies (Section \ref{sec:dynamics}). In Section \ref{sec:discussion} we discuss the limitations in this work and compare to other studies of the CGM, and we conclude in Section \ref{sec:conclusions}.

\section{Simulations and Observational Counterparts}\label{sec:simulations}

\subsection{The Simulations} 
We analyze the FIRE-2 simulations from the FIRE project \citep{Hopkins2018}, a set of cosmological ``zoom-in'' simulations run with the {\sc gizmo} code \citep{Hopkins2015} \footnote{\url{http://www.tapir.caltech.edu/~phopkins/Site/GIZMO.html}} from redshift 99 to 0. The simulations use a Meshless Finite Mass (MFM) hydrodynamic solver with explicitly modelled feedback processes. These processes include energy, momentum, mass, and metal fluxes arising from SNe types I\&II, stellar mass-loss (O-star and AGB), radiation pressure (UV and IR), photo-ionization, and photo-electric heating. Sub-grid metal diffusion is included for four of the ten zoom-in simulations. The simulations explicitly follow chemical abundances of nine metal species (C, N, O, Ne, Mg, Si, S, Ca, and Fe), with enrichment following each source of mass return individually. O and Mg are primarily produced in SNII/Ib/Ic, while C is made in both SNII/Ib/Ic and AGBs, and finally Si is primarily produced in SNII/Ib/Ic and SNIa. The simulations include the cosmic UVB background model of \citet{Faucher2009} together with local radiation sources. Self-shielding is treated with a local Sobolev/Jeans-length approximation. For more information about the simulations and feedback prescriptions, please refer to \citet{Hopkins2018}.

We adopt a standard flat $\Lambda$CDM cosmology with cosmological parameters $H_{0}=70.2 \rm kms^{-1}Mpc^{-1}$, $\Omega _{\Lambda} =0.728$, $\Omega_{m}=1-\Omega_{\Lambda}=0.272$, $\Omega_{b}=0.0455$, $\sigma _{8}=0.807$ and $n=0.961$.

  \begin{table*}
  \begin{threeparttable}
	\caption{Simulated galaxies in this study and their parameters.}
	\label{tab:parameters}
	\begin{tabular}{lccccccccr} 
		\hline
Name & log $M_{halo}^{0}$ & log $M_{stellar}^{0}$ & log $M_{gas}^{0}$&$M_{gas}^{0}/t_{dyn}^{0}$& $SFR ( \rm M_{\rm \odot}/yr)$ & $R_{\rm vir}^{0}$ & $m_{b}$& log $m_{dm}$&Reference\\

   &  ($ \rm M_{\rm \odot}$)&  ($ \rm M_{\rm \odot}$) &($ \rm M_{\rm \odot}$) &($ \rm M_{\rm \odot}/yr$)&$z=0$& (kpc)&($ \rm M_{\rm \odot}$)&($ \rm M_{\rm \odot}$)  &        \\
		\hline
$m11a$ & 10.6 & 8.1 & 9.1 &0.54  & 0.00&  90.3&   2100 &   4.0   & 1\\
$m11b$ & 10.6 & 8.1 &9.3&0.91 &0.02& 92.4 &    2100 &  4.0   & 1\\
$m11c$ & 11.1 & 8.9 &  9.6  &2.00&0.08&  137.0&   2100 &  3.0   & 1\\
$m11d\_md$ & 11.5 & 9.6 & 10.2 &7.52 &0.04&  169.2&  7070 &  4.5 & 2\\
$m11e\_md$ & 11.2 &9.2 &9.7 &2.26&0.32&  135.6&  7070 &  4.5 & 2\\
$m11f$ & 11.7&10.4 & 10.3 &8.87 &6.61&   207.6&  1.7e4 &  4.9  & 1 \\
$m11h\_md$ &11.4 &9.6 &9.8 &2.80&0.45& 146.4 &  7070 &  4.5 &2 \\
$m11i\_md$ &10.9 & 9.0 &9.1 &0.63&0.09&  106.0 &  7070&  4.5 &2\\
$m11q$ & 11.1 & 8.7 & 9.7 &2.24  &0.03&  138.8& 880  & 3.6 &3 \\
$m11v$ & 11.4 & 9.4 &10.2&7.38&0.73&  167.4 &  7070&  4.5 &3 \\ 
		\hline
	\end{tabular}
        \begin{tablenotes}
      \small
      \item     
      parameter units are physical:\\
      (1) Name: Simulation designation, the simulations which ends with 'md' have metal diffusion included. \\
      (2) $M_{halo}^{0}$: Approximate mass of the $z=0$ ``main'' halo (most massive halo in the high-resolution region).\\
      (3) $M_{stellar}^{0}$: Stellar mass of the central galaxy in the main halo at $z=0$.\\
      (4) $M_{gas}^{0}$: Gas mass of the central galaxy in the CGM at $z=0$, the CGM region is the volume beyond the $80\%$ stellar mass radius and within $R_{\rm vir}$ of the main halo. \\
      (5) $t_{dyn}^{0}$: Dynamical time of the main halo at $z=0$, the halo dynamical times of all simulations are between 2.1e9-2.2e9 yr. \\
      (6)  $SFR$: Star formation rate of the galaxy at $z=0$, averaged over the last ten snapshots.\\
      (7) $R_{\rm vir}^{0}$: Virial radius of the $z=0$ ``main" halo, using the definition in \citet{Bryan1998}.\\
      (8) $m_{b}$: Initial gas particle mass\\
      (9) $m_{dm}$: Dark matter particle mass\\
      (10) Reference: Where the simulation is first presented. 1: \cite{Chan2018}, 2: \cite{El-Badry2018}, and 3: \cite{Hopkins2018}.\\
    \end{tablenotes}
\end{threeparttable}
\end{table*}

\subsection{Observational data} 
Since the CGM is diffuse and dim, most observations use absorption spectra of background quasars (QSOs) to gain information about the chemical component along specific sightlines, rather than through direct emission.
Commonly observed absorption lines in the circumgalactic medium include hydrogen \ion{Ly}{$\alpha$} $\lambda 1215$, \ion{Si}{ii} $\lambda 1260$, \ion{Si}{iii} $\lambda 1206$, \ion{C}{ii} $\lambda 1334$, the \ion{C}{iv} $\lambda\lambda$$1548$, $1550$ doublet, the \ion{Mg}{ii} $\lambda\lambda$$2796$, $2803$ doublet, and the \ion{O}{vi} $\lambda\lambda$ $1031$, $1036$ doublet. In this work, we compare our simulations to three observational samples that looked at dwarf galaxies: the COS-Dwarfs survey \citep{Bordoloi2014}, the dwarf galaxies in \citet{Liang2014} and the observations in \citet{Johnson2017}. We describe all three here.

In the COS-Dwarfs survey, 43 low-mass galaxies ($10^{8.2} \rm M_{\rm \odot}<M_{\star}<10^{10.0} \rm M_{\rm \odot}$) over the redshift range $0.010<z <0.104$ were observed using background QSOs. The sources were selected based on galaxy properties, quasar brightnesses and their projected separations. The survey focused on the \ion{C}{iv} 1548 1550 doublet transition in low redshift galaxies, constrained by the spectral coverage of the COS FUV gratings. This transition traces chemically enriched warm gas (${\sim} 10^{4} \rm K $) or highly ionized gas around the host galaxy. They found that the \ion{C}{iv} absorption strength decreases with radius as a power law ($W_r(R) \propto R^{-1}$) and that no absorption was detected beyond $0.5 R_{\rm vir}$. The authors estimate the total carbon mass in the CGM of the galaxies to be comparable to the carbon mass in the interstellar medium of the host galaxy; this conclusion can be directly compared to the simulation data we analyze here. Further, they found a tentative correlation between \ion{C}{iv} absorption strengths and host galaxy sSFR; we address this and similar correlations in section \ref{sec:global_properties}.

\citet{Liang2014} studied absorption transitions of \ion{Ly}{$\alpha$}, \ion{C}{ii}, \ion{C}{iv}, \ion{Si}{ii}, \ion{Si}{iii}, and \ion{Si}{iv} in the vicinities of 197 galaxies. They identified their sample from a comparison of 300 spectroscopic galaxies in the public archives, with background QSOs having high-quality spectra from the Hubble Space Telescope (HST) data archive. The sample has a median redshift of $\langle z\rangle=0.176$, a median impact parameter of $\langle d\rangle=362$ kpc and a median stellar mass of $log (\rm M_{star}/M_{\rm \odot})=9.7 \pm 1.1$. From $d<0.3R_{\rm vir}$ to $d {\sim} 0.7R_{\rm vir}$, a differential covering fraction between low- and high-ionization gas is measured, suggesting that the CGM becomes progressively more ionized with increasing galactocentric radius; At $d \gtrsim 0.7R_{\rm vir}$ no metals in either the low- or high-ionization states are detected from both the low-mass dwarfs and high-mass galaxies, though abundant hydrogen gas is observed to well beyond $R_{\rm vir}$, with a mean covering fraction of ${\sim} 60$ per cent.

\citet{Johnson2017} analyzed the CGM of 18 star-forming field dwarf galaxies ( $log (\rm M_{star}/M_{\rm \odot})= 7.7-9.2$ ) at $z=0.09-0.3$, drawn from a Magellan telescope survey in the fields of quasars with high S/N COS absorption spectra. The authors searched for \ion{H}{i}, \ion{Si}{ii}, \ion{Si}{iii}, \ion{Si}{iv} and \ion{O}{vi} absorption in the quasar spectra. They found that compared to massive star-forming galaxies, metal absorption is less common from the CGM of dwarf galaxies, and is weaker with the exception of \ion{O}{vi}.

In this paper, we will compare the structure of the CGM around low-redshift dwarf galaxies inferred by the FIRE-2 simulations with the results of the available observations \citep{Bordoloi2014,Liang2014, Johnson2017}. For reference, the minimum detected EWs for \ion{C}{iv} are 95 $m\AA$, 59 $m\AA$, and 130 $m\AA$ respectively; for \ion{Si}{III} the minimum EWs are 54 $m\AA$, 70 $m\AA$ for the later two references; for \ion{O}{vi} the minimum detected EW is 80 $m\AA$ (only \citet{Johnson2017} reports this ion); there are no measurements for \ion{Mg}{II}.

\section{Generating absorption spectra from simulation data} \label{sec:post-processing}

To compare the results of the FIRE simulations and the observational data, we must select simulated galaxies that match the mass and redshift properties of the observational studies. Further, we must model a representative sample of spectra from sightlines using the simulated properties, including the density, temperature, and metallicity of the gaseous medium in the selected galaxies. 

In this section we describe the selection criteria for galaxies simulated within FIRE. We also describe how absorption spectra have been generated from the simulation data, enabling direct comparison to observational results.

\subsection{Selection of simulated galaxies}
\label{sec:snapshot_selection}

To compare against the available observations, we selected the main galaxy from 10 zoom-in simulations of the FIRE-2 project with stellar masses between $10^{8}\, \rm M_{\sun} < M_{\rm *} < 10^{11}\, \rm M_{\sun}$, representing the stellar mass range of observed dwarf galaxies. The parameters of the selected simulations and their primary halos are listed in \autoref{tab:parameters}. 

For each of the simulations, we selected time-based snapshots that are within the redshift range of the observations ($0 \leq z \leq 0.3$). We find the center of the main galaxy in each snapshot using the Amiga Halo Finder \citep[AHF;][]{Gill2004, Knollmann2009}. We limit ourselves to snapshots that are separated, at a minimum, by the dynamical time of the galaxy's disk; the typical dynamical time of the simulated galaxies is ${\sim} 50$ Myr. We do this to ensure each galaxy snapshot is dynamically independent, and consequently not biased towards artifacts from the simulation or specific configuration of the galaxies. The dynamical time of the halo is an order of magnitude larger than the dynamical time of the galaxy, but the disk dynamical time is more relevant to the small scale structures we seek to probe. We note that we do not place any constraint on the star formation properties or metallicity of the simulated galaxies.

Most of the runs we studied did not include any sub-grid metal diffusion. However, four of the simulations did include a sub-grid model for metal diffusion; these are marked with `\_md' in \autoref{tab:parameters}. The FIRE simulations use a finite mass method which by definition prevents transport of material from one particle to another. In reality, there will be transport of material from one region to another on the particle separation scale. The metal diffusion runs approximate this diffusion using the \citet{Smagorinsky1963} model. We show in Section \ref{sec:EW} that including the sub-grid model for metal diffusion reduced the spread in the EW distribution. 

\begin{figure}
	\includegraphics[width=\columnwidth]{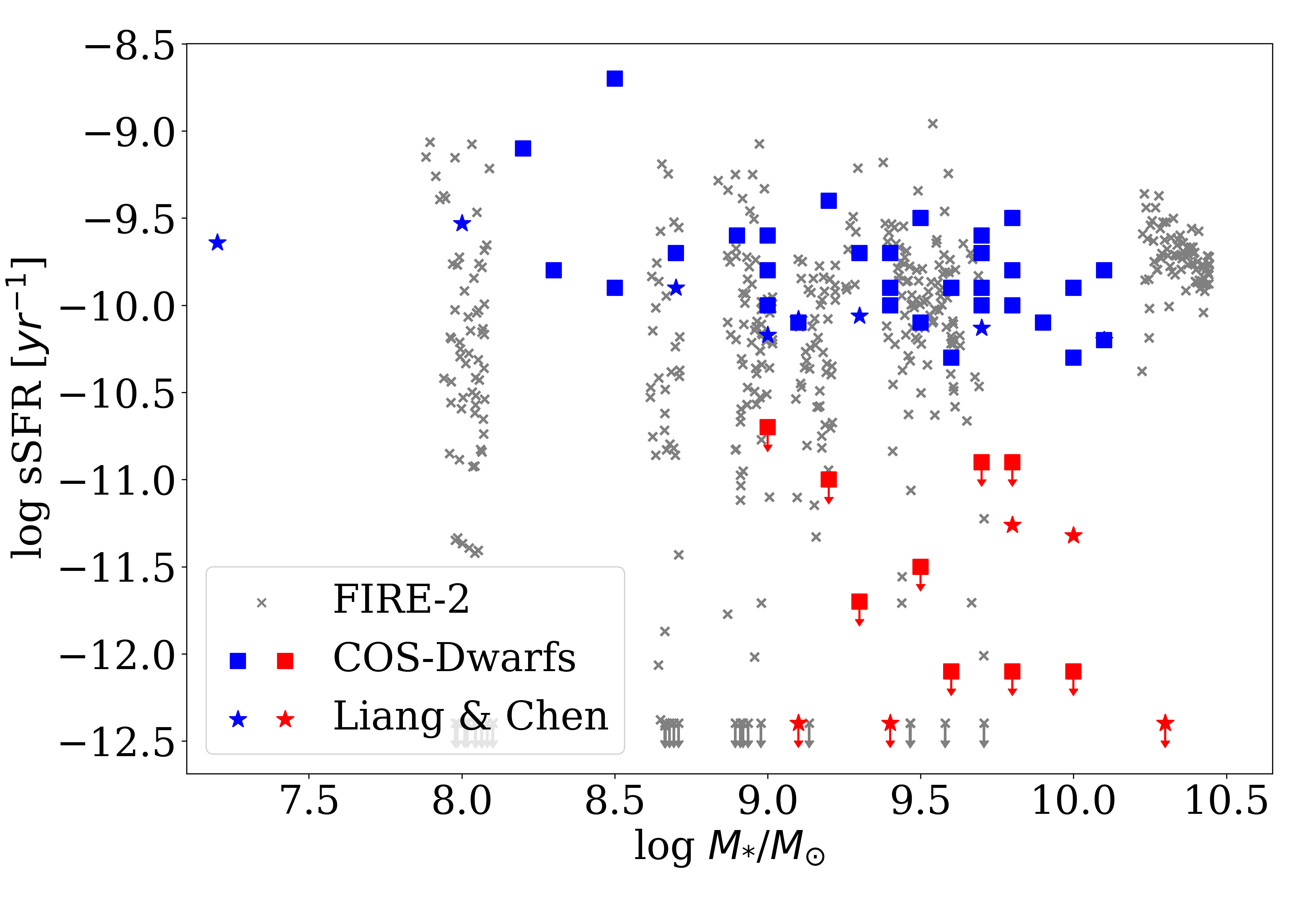}
    \caption{The relation between sSFR and stellar mass in simulation and observations. The grey x's represent dynamically independent simulation snapshots from the FIRE project, where the SFR is calculated by averaging young star particle mass over a 5-Myr timescale, the grey upper limits have sSFR lower than $10^{-12} \rm M_{\rm \odot} yr^{-1}$; the blue and red squares are star-forming and quiescent galaxies from the COS-Dwarfs survey; the blue and red stars are star-forming and passive galaxies from the \citet{Liang2014} study. SFGs and passive galaxies are separated at sSFR=$10^{-10.6} \rm M_{\rm \odot} yr^{-1}$. The \citet{Johnson2017} sample is not included in this figure because of the unknown sSFR of the galaxies.}
    \label{fig:sSFR_stellar_mass}
\end{figure}

These selection criteria resulted in ${\sim} 40$ snapshots for each galaxy with redshift $\leq 0.3$, leading to 442 samples in total. The mass and specific star formation rate (sSFR) of each sample galaxy are plotted in \autoref{fig:sSFR_stellar_mass}, alongside the properties of observed galaxies from the COS-Dwarfs survey, and \citet{Liang2014}. In order to study the star formation rate dependence of the CGM properties, we subdivide the sample into star-forming galaxies and quiescent galaxies, using the sSFR criterion of $10^{-10.6}$ $ \rm yr^{-1}$, as done for COS-Dwarfs. The simulated sample galaxies match the stellar mass range from the observations. The distribution of sSFR for $M_*>10^9 M_\odot$ matches that of the observational sample. However, for $10^8<M_*/M_\odot<10^9$, the simulations predict a large number of quiescent galaxies not found in the samples we compare to. 

\subsection{Generating Synthetic Spectra}

For each sample galaxy, we choose ${\sim} 135$ random lines of sight (LOS), sampling the space uniformly (equal number per unit projected area). We limit ourselves to impact parameters between $0 \leq b \leq 1.3R_{\rm vir}$. We use the definition of $R_{\rm vir}$ in \citet{Bryan1998}, consistent with the observations. 

The length of the LOS is $6R_{\rm vir}$, so that the result will not be strongly affected by the low-resolution IGM region of the simulation. At a distance of $3R_{\rm vir}$, the density in the simulations has dropped to $n\approx 2\times10^{-7}cm^{-3}$ (roughly $\Omega_b n_{crit}$ where $\Omega_b\approx 0.04$ is the cosmic baryon fraction and $n_{crit}=\rho_{crit}/m_p$, with $\rho_{crit}$ the cosmic mean density), while the metallicity is $[M/H]\approx-3$ (see section \ref{sec:phys_dist}), comparable to the mean metallicity of the IGM $[O/H]=-2.85$ \citet{Simcoe2004}. The maximum path length is set by the velocity offset $\Delta v$ beyond which the absorber is no longer taken to be associated with the target galaxy. Following \citet{Bordoloi2014} we take  $\Delta v=600km\,s^{-1}$, which corresponds to a path length $D=\Delta v/ H_0\approx8 Mpc$. Taking the fraction of $O_{\rm VI}=0.2$, we find $N_{O\,VI}\approx1.6\times10^{12}cm^{-2}$. As we will show later, this is about $20\%$ of the lowest column we find for impact parameters $b<R_{\rm Vir}$. Our estimates are biased low by this amount compared to observations.

A direct comparison to the observational work requires the generation of spectral equivalent widths in each of the desired absorption lines. To do this, we need to estimate the density of the relevant ions. The FIRE simulation tracks 11 different atoms (H, He, C, N, O, Ne, Mg, Si, S, Ca and Fe). However, performing full radiative transfer to determine the ionic states would be computationally expensive to run at the desired resolution. 

We use the open-source code Trident \citep{Hummels2017Trident} to post-process the FIRE simulations for analysis and comparison with the observational sample. We apply Trident to infer the abundance of four ions throughout the simulations: \ion{C}{iv}, \ion{O}{VI}, \ion{Mg}{ii}, and \ion{Si}{iii} accounting for photoionization and collisional ionization. Trident estimates these abundances using a detailed look-up table assembled from over $10^6$ Cloudy \citep{Ferland2013} radiative transfer simulations, systematically varying density, temperature, and a redshift-dependent radiation field consistent with the Haardt-Madau UV/X-ray background \citep{Haardt2012}.  

The effects of self-shielding are well known to play a role in regions of high gas density, where the abundance of some low-ionization-energy ions are partially shielded from ionizing radiation \citep{Rahmati2013MNRAS.430.2427R}. We account for this effect using new updated Trident ion tables using a method first described in \citet{Emerick2019MNRAS.482.1304E} for the Grackle cooling code \citep{Smith2017}. These tables are constructed as described above from Cloudy simulations, each initialized with a single density, temperature, and radiation field. Each Cloudy run is a one dimensional simulation with length equal to the Jeans length but capped at 100 pc for large Jeans lengths. The radiation field is applied to one end of the simulation volume, and the simulation executes until it reaches an equilibrium state. At the end of this period, the ionic abundances are measured at the other end of the simulation volume, self-consistently capturing the effects of self-shielding by how effectively radiation is able to penetrate the intervening gas on the maximum scale it would clump.

Using these post-processed ionic abundances added to our FIRE galaxies, we generate ${\sim} 135$ random sightlines through the each simulation volume at a variety of different impact parameters relative to the center of each galaxy. These sightlines are meant to mimic the observational datasets, each representing the path of light from a distant quasar through the intervening CGM of our foreground galaxies. The gas properties and ionic abundances for each sightline are passed to Trident \citep{Hummels2017Trident} to generate synthetic spectra, depositing absorption features in the spectra for each absorbing ion encountered along the line of sight. Finally, we numerically integrate the absorption features for each ionic transition to identify an equivalent width (EW) value appropriate for direct comparison with the observational samples. We have visually inspected the spectra of \ion{Mg}{II}, and almost all of the lines are optically thin (the double ratio is 2:1); only 1 LOS out of 6000 had a ratio that was clearly lower. 

\section{Results}\label{sec:results}

\subsection{Ion Equivalent Width vs Impact Parameter}\label{sec:EW}

\begin{figure*}
	\includegraphics[width=\textwidth]{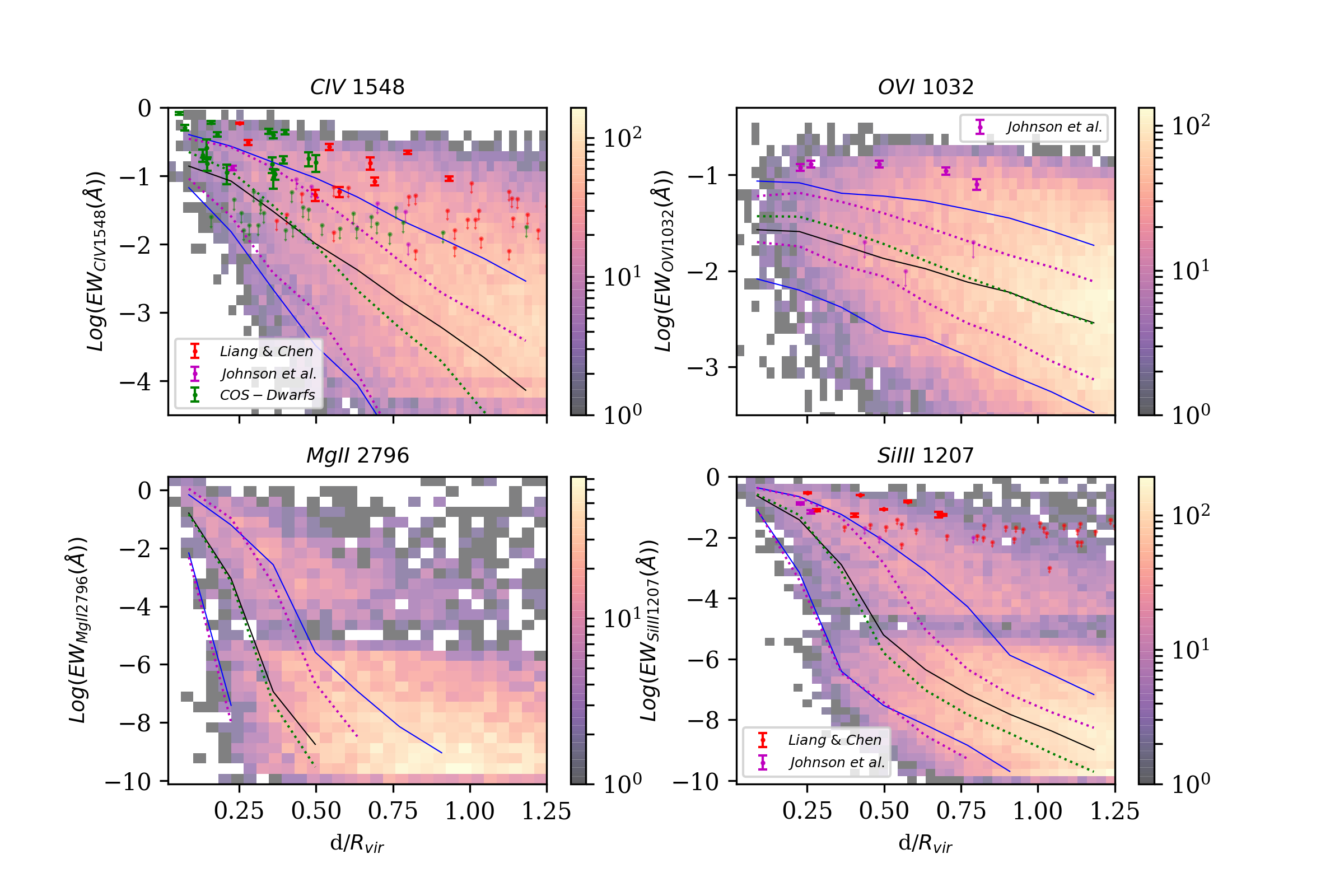}
    \caption{Rest-frame absorption EW of four ions (\ion{C}{iv}, \ion{O}{VI}, \ion{Mg}{ii} and \ion{Si}{iii}) versus normalized impact parameter. The 2-D histograms of EW of mock LOS from all dynamically-independent simulation snapshots (defined in Section \ref{sec:snapshot_selection}) are shown in the background, with the solid black lines showing the median values and the solid blue lines showing the 16-84 per cent values of the 6 runs without metal diffusion (\autoref{tab:parameters}), and the dotted lines showing the statistics for the 4 runs with metal diffusion (m11d, m11e, m11h, and m11i). Available observational data are included for each ion, and the data points with only downward arrows are $2\sigma$ upper limits on non-detections. The amount of \ion{C}{iv} predicted by the FIRE simulations is roughly consistent with observational constraints. The equivalent widths of the other ions inside ${\sim} 0.5 R_{\rm vir}$ are underpredicted by a factor of 3 to ten. Beyond half a virial radius, there are only non-detections from observations.}
    \label{fig:EW}
\end{figure*}

We compare the generated absorption spectra of the four ions in the CGM of the simulation sample to the observations using the measure of EW. In this subsection, we compare the distributions of absorption EW as a function of impact parameter, which is directly measured in observations.

We show the rest-frame equivalent widths of four ions (\ion{C}{iv}, \ion{O}{VI}, \ion{Mg}{ii} and \ion{Si}{iii}) as a function of normalized impact parameter in \autoref{fig:EW}, and compare to observational data from \citet{Bordoloi2014}, \citet{Liang2014}, and \citet{Johnson2017}. The EW from all mock LOS from all 384 dynamically-independent snapshots are shown as 2-D histograms in the background in \autoref{fig:EW}, with the black and blue lines showing the median and 16-84 per cent (${\sim}\pm$ 1 $\sigma$) values, and available observational data are included in the panels for each ion. 

\ion{C}{IV} has the highest number of observations among the four ions. Within $0.5R_{ \rm vir}$, $50 \pm 12 \%$ of the observations are above the detection limit, and $27 \%$ of the simulated sightlines are above the detection limit (${\sim}$100 \AA), so the simulated EW -- impact parameter distribution for \ion{C}{iv} is somewhat low compared to observations. At large radii, the simulation underpredicts the EW by a large factor. 

For \ion{O}{VI}, there are five detections and three upper limits available from observations. The simulated \ion{O}{VI} EW distribution is about a factor of three lower than the observed distribution, since the detections are all above the 84th percentile in the simulations and the upper limits are around the simulated median. Although this could be due to the fact that the galaxies in \citet{Johnson2017} are all star-forming (without available estimates of SFR), and as discussed in Section \ref{sec:global_properties}, \ion{O}{VI} EW increases with SFR. 

For \ion{Si}{iii}, there are 10 detections available, all at $d/R_{\rm vir} < 0.75$. The detections fall above the predicted median of the simulated EW distribution. 

For \ion{Mg}{II}, there is no observational data available for dwarf galaxies to compare against; the prediction from the simulations is that it will be very difficult to detect this ion in absorption.

In general, the EWs of the four ions all decrease with normalized impact parameter ($d/R_{\rm vir}$) -- with shallower fall-off rates for \ion{C}{IV} and \ion{O}{VI}, and steeper fall-off rates for \ion{Si}{iii} and \ion{Mg}{II}, indicating that there is little \ion{Mg}{II} and \ion{Si}{III} in the outer CGM of dwarf galaxies (see Section \ref{sec:phys_dist}). In the case of \ion{Si}{III}, it appears that the EW falls off more rapidly with impact parameter in the simulations than it does in the observations. And for all ions, there are wide ranges of possible EW values at high impact parameters. For the two low ionization potential ions, there are significant fractions of LOS with EW consistent with zero 0 \AA\, in the simulations: 96.8\% for \ion{Mg}{II}, and 76.0\% for \ion{Si}{III}.

There are gaps in the EW distributions of \ion{C}{IV} (around $10^{-4.4}$\AA), \ion{Mg}{II} (around $10^{-5.2}$\AA), and \ion{Si}{III} (around $10^{-5.0}$\AA) in \autoref{fig:EW}, which are more clearly seen in \autoref{fig:corner_plot_ionew}. The gaps are due to the fact that most CGM gas reside either on the low temperature or high temperature side of the peak in the cooling function, and not at the peak. This is discussed further in the next section.

\subsection{CGM Phase Space}\label{sec:CGM_phase_space}
\begin{figure*}
	\includegraphics[width=\textwidth]{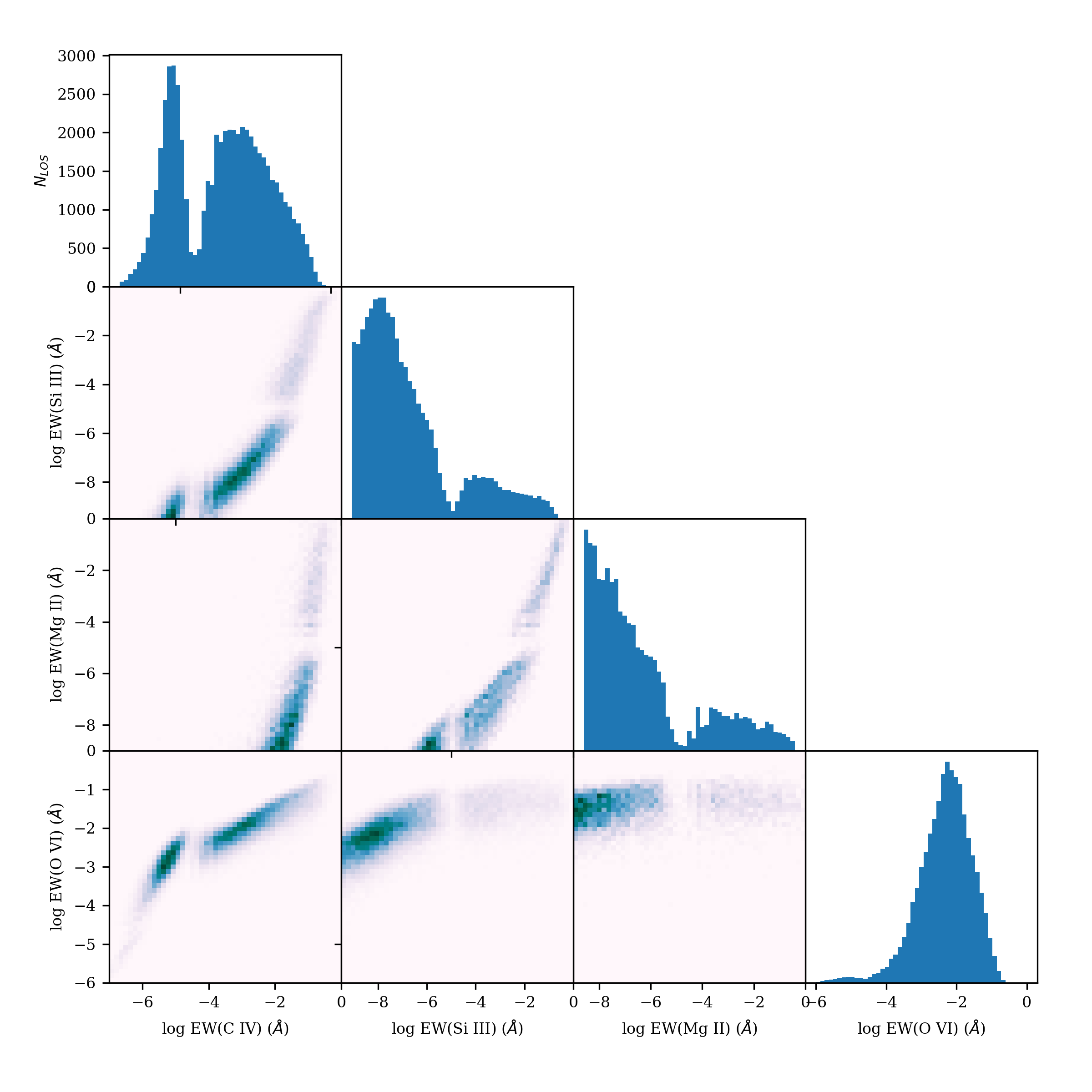}
    \caption{Distributions of the EW of the four ions in this study along the same LOS. The plots along the diagonal show histograms of the EWs of the four ions. The \ion{C}{IV}, \ion{Si}{III}, and \ion{Mg}{II} histograms have more than one peak due to atomic physics: the CGM gas associated with the two peaks in the EW histograms reside on the low $T$ and high $T$ side of the peak in the cooling function, leaving a gap (corresponding the the maximum cooling rate) in between (see Section \ref{sec:CGM_phase_space}). The off-diagonal panels display the relationships between the EWs of the ions along the same LOS. Note that only 3.2\% and 24\% of LOS for \ion{Mg}{II} and \ion{Si}{III} are included in this figure, since the rest of the LOS have $EW \approx 0$ \AA\,for the two ions; and not all LOS shown in the histograms appear in the off-diagonal panels, because phases of the CGM gas probed by the different ions do not completely overlap with each other. There are strong correlations between some pairs of ions, such as \ion{O}{VI} and \ion{C}{IV}, \ion{Si}{III} and \ion{C}{IV}, and \ion{Mg}{II} and \ion{Si}{III}; and no correlation between \ion{O}{VI} and \ion{Mg}{II}.}
    \label{fig:corner_plot_ionew}
\end{figure*}

Having generated the absorption spectra of \ion{C}{IV}, \ion{O}{VI}, \ion{Mg}{II}, and \ion{Si}{III} along identical LOS, we can use them to examine the relationship between the EWs of the ions in the simulations. \autoref{fig:corner_plot_ionew} shows the EW distributions of the four ions from all sightlines produced, and their mutual relationships. 

The EW histograms of \ion{C}{IV}, \ion{Si}{III}, and \ion{Mg}{II} in \autoref{fig:corner_plot_ionew} display bimodality; in \ion{C}{IV} the gap is at $EW{\sim} 10^{-4.4}$\AA, in \ion{Si}{III} the gap is at ${\sim} 10^{-5.0}$\AA, in \ion{Mg}{II} the gap is at ${\sim} 10^{-5.2}$\AA. We do not see a gap in \ion{O}{VI}. The existence of the gaps in the histograms is connected with the multi-modal distributions of the ion temperatures: most of the gas is around $T {\sim} 10^{5}K$, which is likely shock heated and produces the larger EW peaks with higher EW values and higher column densities; The LOS in the lower EW peaks have lower column densities for the corresponding ion compared to the LOS in the higher EW peak, and is associated with $T {\sim} 10^{4}K$ gas, which is likely photoionized. This is the effect of the cooling curve which produces a multiphase medium that imprints itself as discrete temperatures and correspondingly discrete ionization fractions per particle. These create the observed multimodalities in the EW distributions. The reason we do not see multimodality in \ion{O}{VI} is that the low temperature gas is insufficiently energetic to reach that ionization state.

The off-diagonal plots in \autoref{fig:corner_plot_ionew} show relationships between the EWs of the four ions. There are correlations between some pairs of ions, such as \ion{O}{VI} and \ion{C}{IV}, \ion{Si}{III} and \ion{C}{IV}, and \ion{Mg}{II} and \ion{Si}{III}; and little correlation between \ion{O}{VI} and \ion{Mg}{II}, or between \ion{Mg}{II} and \ion{C}{IV}. The correlations indicate which pairs of ions probe the same phase of the CGM gas, and this is largely consistent with the difference between their respective ionization energies. For example, in the case of \ion{Mg}{II} (ionization energy of 15 eV) and \ion{O}{VI} (ionization energy of 138 eV), any given LOS containing absorption from \ion{Mg}{II} will also have absorption from \ion{O}{VI}. Since \ion{O}{VI} is located at large galactocentric radii, and \ion{Mg}{II} is located at small galactocentric radii, there is little correlation between the two ions. Note that the \ion{O}{VI} is high in EW for all LOS with \ion{Mg}{II}, because of the large path length through the halo at small impact parameters, the latter being the only LOS where the simulations predict significant amounts of \ion{Mg}{II}. We discuss this further in Section \ref{sec:phys_dist}.

\subsection{Connecting UV Absorption to Global Galactic Properties} \label{sec:global_properties}

\begin{figure*}
	\includegraphics[width=\textwidth]{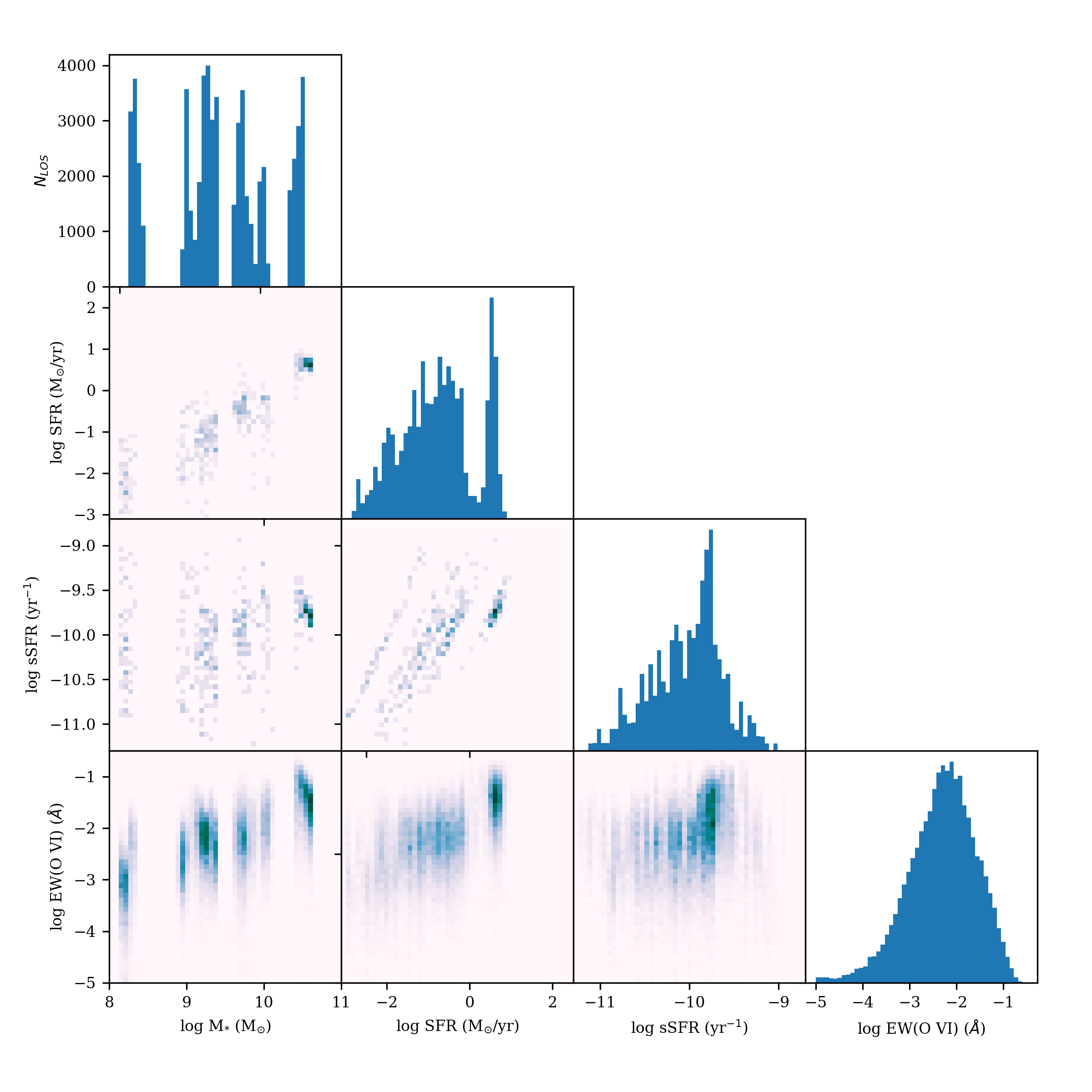}
    \caption{Distributions of \ion{O}{VI} EW and three galactic properties. The diagonal plots show histograms of four quantities -- $M_{\rm *}$, SFR, sSFR, and \ion{O}{VI} EW. The off-diagonal plots display the relationships between all of these properties. In particular, the bottom off-diagonal panels show correlations between \ion{O}{VI} EW with $M_{\rm *}$, SFR, and sSFR (correlation coefficients presented in \autoref{tab:cc}). The EW of the other ions in this study (\ion{C}{IV}, \ion{Mg}{II}, and \ion{Si}{III}) show similar trends with the four galactic properties. Other strong correlations with CGM absorption include $M_{\rm H}$, $M_{\rm *}/M_{\rm H}$, and $R_{\rm vir}$. The ions in our analysis do not show strong correlation with sSFR (as shown in the bottom panel in the third column), impact parameter, or redshift.}
    \label{fig:corner_plot}
\end{figure*}

\begin{table}
  \begin{threeparttable}
	\caption[Table]{Correlation coefficients of EW, ordered by decreasing ionization potential, and various global galactic quantities. The errors of the correlation coefficients are ${\sim} 0.004$, calculated using the bootstrap sampling method. The correlation strength increases with the ionization potential of the ion.}\label{tab:cc}
	\begin{tabular}{lcccr} 
		\hline
  & \ion{O}{VI}& \ion{C}{iv}  & \ion{Si}{iii} &   \ion{Mg}{ii} \\
		\hline
$M_{\rm *}$ & 0.58&0.39  &0.28  &0.25 \\
$M_{\rm H}$ & 0.53& 0.35  & 0.25  &0.23 \\
$M_{\rm *}/M_{\rm H}$ &0.57& 0.37 &0.25  & 0.23\\
$R_{\rm vir}$  &0.55&0.36  &0.25  & 0.24\\
SFR & 0.51& 0.36 & 0.26 &0.23 \\
sSFR &0.29& 0.21  & 0.15 &0.15 \\
$d$  & 0.15& -0.05 & -0.16 & -0.13\\
$d/R_{\rm vir}$  &-0.31& -0.41 &-0.46  &-0.37 \\
$z$  & 0.00& -0.02 & -0.05 & 0.02\\ 
		\hline
	\end{tabular}
  \end{threeparttable}
\end{table}

We investigate the relationship between UV absorption in the CGM and a series of global galactic properties. The global properties we examine are $M_{\rm *}$, $M_{\rm H}$, $M_{\rm *}$/$M_{\rm H}$, SFR, $R_{\rm vir}$, and sSFR; the other properties we study are redshift, impact parameter d, and $d/R_{\rm vir}$.

We find that all of the global properties correlate with one another, except for sSFR; and CGM absorption lines correlate with all of the global property correlations, again, except for sSFR. \autoref{fig:corner_plot} shows the distributions of \ion{O}{VI} EW and three selected galactic properties. The bottom off-diagonal panels illustrate that there are strong correlations between \ion{O}{VI} EW and two global galactic properties -- $M_{\rm *}$ and SFR; in contrast, there is not any significant trend of \ion{O}{VI} EW with sSFR. This is also true for the other ions (\ion{C}{IV}, \ion{Mg}{II}, and \ion{Si}{III}) analyzed in this work, although with weaker correlation strengths with decreasing ion ionization potential. In general, The absorption strengths of the ions increase with $M_{\rm *}$, $M_{\rm H}$, $M_{\rm *}/M_{\rm H}$, SFR, and $R_{\rm vir}$, and decrease with $d/R_{\rm vir}$, with the correlation strengths increasing with ion ionization potential. The correlation coefficients between EW and various properties are listed in \autoref{tab:cc}.

The CGM absorption has a weak dependence on redshift, sSFR and impact parameter in our analysis. The weak dependence on redshift is not surprising considering that the simulation data in our analysis is in the narrow redshift range of 0 - 0.3. However, it is reasonable to expect that the strength of absorption lines to increase with sSFR -- when fixing the $M_{\rm *}$ of a galaxy, the EW of ions could increase with SFR if the material in the halo is due to star formation related winds. So the lack of correlation of ion absorption with host galaxy sSFR is at first glance puzzling. We checked the variation of SFR over time, and it appears to occur in short bursts \citep{Muratov2015, Angles2017}, so the material in the halo is not due to steady winds. It seems likely that the material in the halo was put in place by earlier star bursts. This is consistent with a particle tracking analysis \citep[][more on this in Section \ref{sec:Origin}]{Hafen2019}. Consequently, on closer examination, the lack of correlation between sSFR and material in the halo is not that surprising.

\begin{figure}
    \includegraphics[width=\columnwidth]{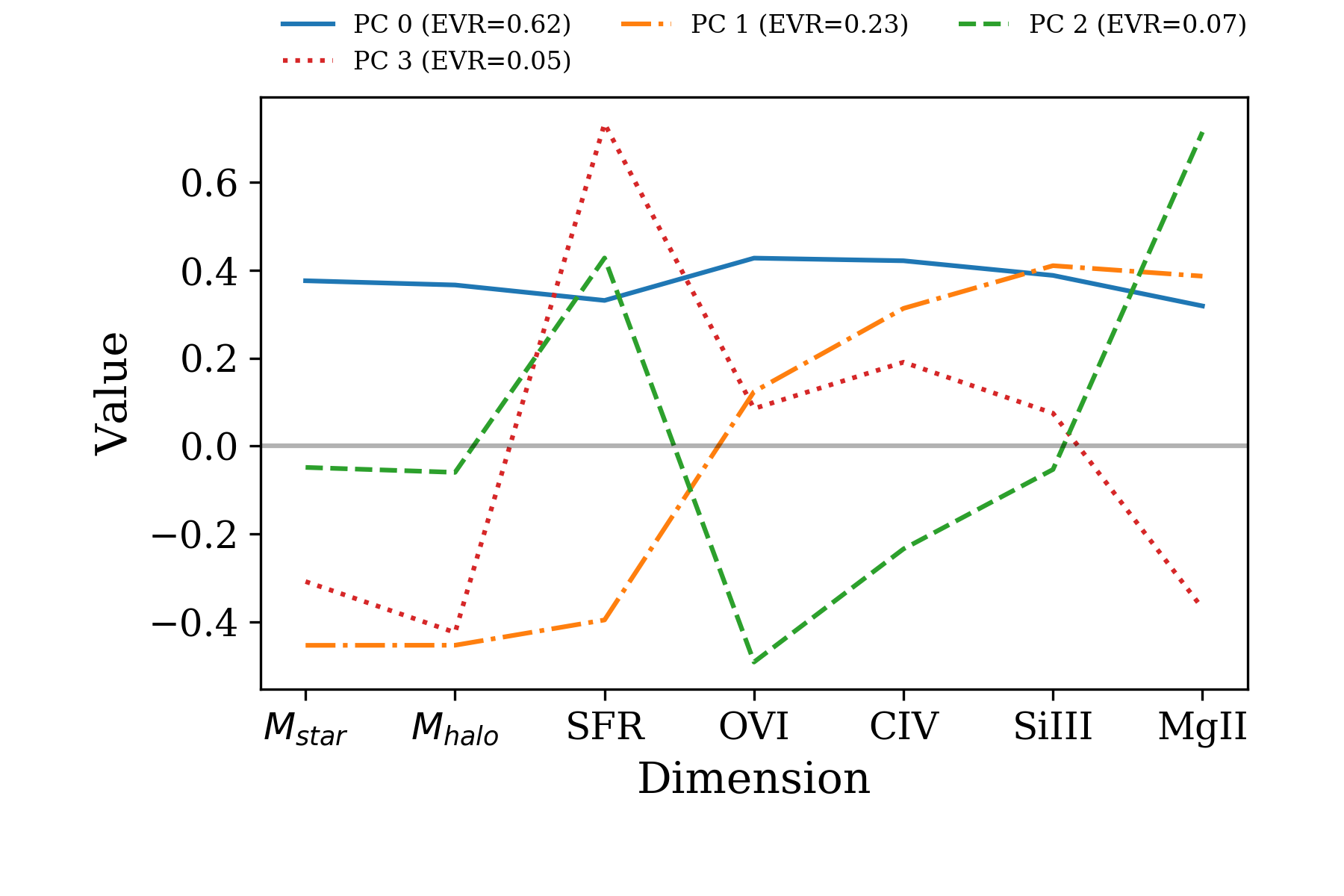}
    \caption{The four principle components which account for 98.2\% of the variance in the original data. The features involving absorption lines on the x axis represent the absorption EWs of the ions, and the values of all dimensions are in log space except for SFR. The features have been transformed so that there is zero mean and unit variance. The percentage of variance preserved in each principle component is shown in the legend. 62 \% of the variance is accounted for by a single component, which we interpret as a correlation between halo mass and all other forms of mass, including stellar mass and gas mass in the CGM. The second component, which accounts for 23 \%, we interpret as a temperature effect. Lower mass halos have lower virial temperatures, which result in higher absorption for the low-ionization potential ions \ion{Mg}{ii} and \ion{Si}{iii}.}
    \label{fig:PCA}
\end{figure}

To further study the relationship between the global properties and the absorption strengths of the ions, we perform a principal component analysis on the data. \autoref{fig:PCA} shows the first four principle components which capture 98.2\% of the variance of the data when projected into a 7-dimensional space. The first principle component (labeled `PC 0' on the figure), which accounts for 62\% of the variance, shows that the three global galaxy properties (halo mass, stellar mass, and star formation rate) and the ion absorption strengths are all correlated. This component may be interpreted as indicating that more massive halos (and hence more massive galaxies) contain more gas, which in turn produce stronger ionic absorption. The second principle component (`PC 1'), which accounts for 23\% of the variance, shows that the absorption strengths are anti-correlated with all three global galaxy properties, with lower ionization state ions being more strongly anti-correlated with the galaxy properties. We interpret this component as an indicator of the temperature of the CGM; higher mass halos have higher virial temperatures, hence fewer low ionization ions compared to lower mass (and temperature) halos. The third principle component (`PC 2') shows that while the absorption strength of three ions are anti-correlated with SFR, \ion{Mg}{II} is correlated with SFR. This component, which accounts for 7\%, may be an indicator of the stronger winds from feedback associated with star formation. 

\subsection{Physical Distribution of the CGM gas}\label{sec:physical_dist}

\begin{figure*}
	\includegraphics[width=\textwidth]{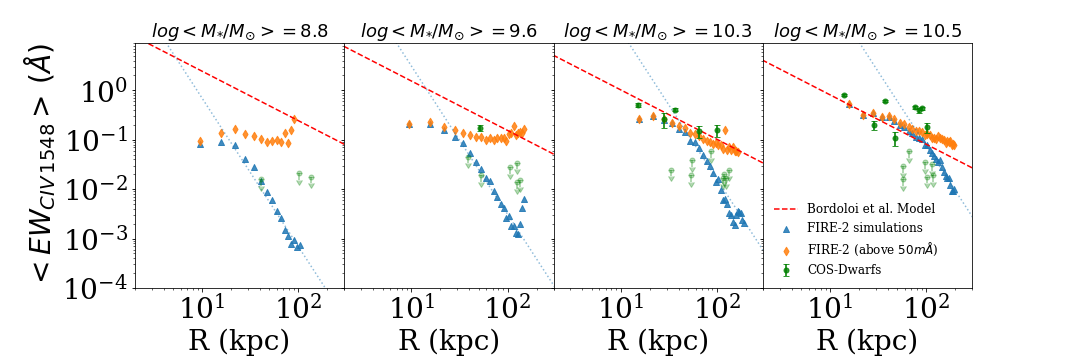}
    \caption{A comparison of the average \ion{C}{iv} absorption EW profile in simulations with the best-fit model from \citet{Bordoloi2014}. The four panels correspond to four $R_{\rm vir}$ bins (with mean $R_{\rm vir}$ values of 100 kpc, 148 kpc, 191 kpc, and 215 kpc), which translate to mean $M_{\rm *}$ values of $6.2\times10^8 \rm M_{\rm \odot}$, $3.9\times10^9 M_{\rm \odot}$, $2.0\times10^{10} \rm M_{\rm \odot}$, and $3.3\times10^{10} M_{\rm \odot}$. In each panel, the blue triangles show the average value of \ion{C}{iv} EW in radial bins in the simulations while the orange diamonds show the average \ion{C}{iv} EW placing a lower detection limit of 50\,m\AA\, similar to the limits in \citet{Bordoloi2014}; the green points are detections and upper limits in the COS-DWARFS survey \citep{Bordoloi2014}; the red dashed line is the best-fit model by \citet{Bordoloi2014}; and the blue dotted line shows a reference $R^{-3}$ profile.
    We find that the simulated \ion{C}{iv} EW profile follows a $R^{-3}$ power law beyond the inner core regions. The \ion{C}{iv} EW profile for sightlines above 50\,m\AA\,in simulations agree well with COS-Dwarf observations and the best-fit model for the two largest $R_{\rm vir}$ bins, but diverge for the smallest halos. The difference between the full and detection-limited mean EW profiles suggests that the form of the best-fit model from \citet{Bordoloi2014} is primarily due to the detection limit bias of the observations.}
    \label{fig:civ_spatial}
\end{figure*}



We investigate the geometry of the ionized gas probed by the absorption lines simulated, and the physics that underlies its distribution.

\subsubsection{Spatial distribution of \ion{C}{iv} absorption} 

We compare the 1-D radial distribution of \ion{C}{IV} absorption EWs in our sample to the proposed profile in \citet{Bordoloi2014}:
\begin{equation} \label{eq: radial}
    \langle W_{r}(R) \rangle = W_{0} \times 
    \Big(\frac{R}{R_x} \Big)^{-1} \times exp \Big(\frac{-R_{\rm vir}}{R_x}\Big)
\end{equation}
where $W_0$ and $R_{x}$ are two free parameters describing absorption and radial scaling. The best-fit values obtained by \citet{Bordoloi2014} from a maximum likelihood fit are $R_x=105 \pm 27 \, kpc$ and $W_0=0.6 \pm 0.3\,$\AA.

In \autoref{fig:civ_spatial}, we divide our sample into 4 $R_{\rm vir}$ bins, roughly equivalent to halo mass, which enables direct comparison with \citet{Bordoloi2014}. We then bin our data radially, and calculate the average \ion{C}{IV} absorption EW in each bin. To account for the effect of the detection limit in the COS-Dwarfs Survey, we repeat the calculation with a detection limit of 50\,m\AA, the lowest detection limit in the COS-Dwarfs survey. The resulting radial profiles are plotted in \autoref{fig:civ_spatial}, along with the observational data and the proposed profile from \citet{Bordoloi2014}. 
    
The \ion{C}{IV} absorption EW profiles derived from our simulations do not agree with the proposed profile in \citet{Bordoloi2014} in slope, parameterization, and trend with $R_{\rm vir}$: the FIRE profiles fall off more steeply with radii (the simulated \ion{C}{iv} EW profile follows a $R^{-3}$ power law beyond the inner core regions), which means that the \ion{C}{IV} ion is more concentrated towards the centre of the halo in the simulations; \autoref{eq: radial} suggests that \ion{C}{IV} absorption EW decreases with $R_{\rm vir}$, while \ion{C}{IV} absorption EW increases with $R_{\rm vir}$ in the simulated sample, and the correlation coefficient between \ion{C}{IV} absorption EW and $R_{\rm vir}$ in the simulated sample is 0.36. 

The conflict between the simulated profile of the \ion{C}{IV} EW and the  modelling in \citet{Bordoloi2014} is, we suggest, due to the detection limit bias. The proposed profile in \citet{Bordoloi2014} is based on a limited sample with 17 detections in total, with no detections beyond 100 kpc, and the two lower $R_{\rm vir}$ bins are not well constrained with only one detection. The fact that the \ion{C}{iv} EW profile for sightlines above 50\,m\AA\,in simulations agree well with both the COS-Dwarf observations and the best-fit model from \citet{Bordoloi2014} for the two largest $R_{\rm vir}$ bins suggests that this inferred profile is due to the detection limit bias in observations.

\subsubsection{Physical Distribution of Ions and Metals in the CGM} \label{sec:phys_dist}

\begin{figure*}
	\includegraphics[width=\textwidth]{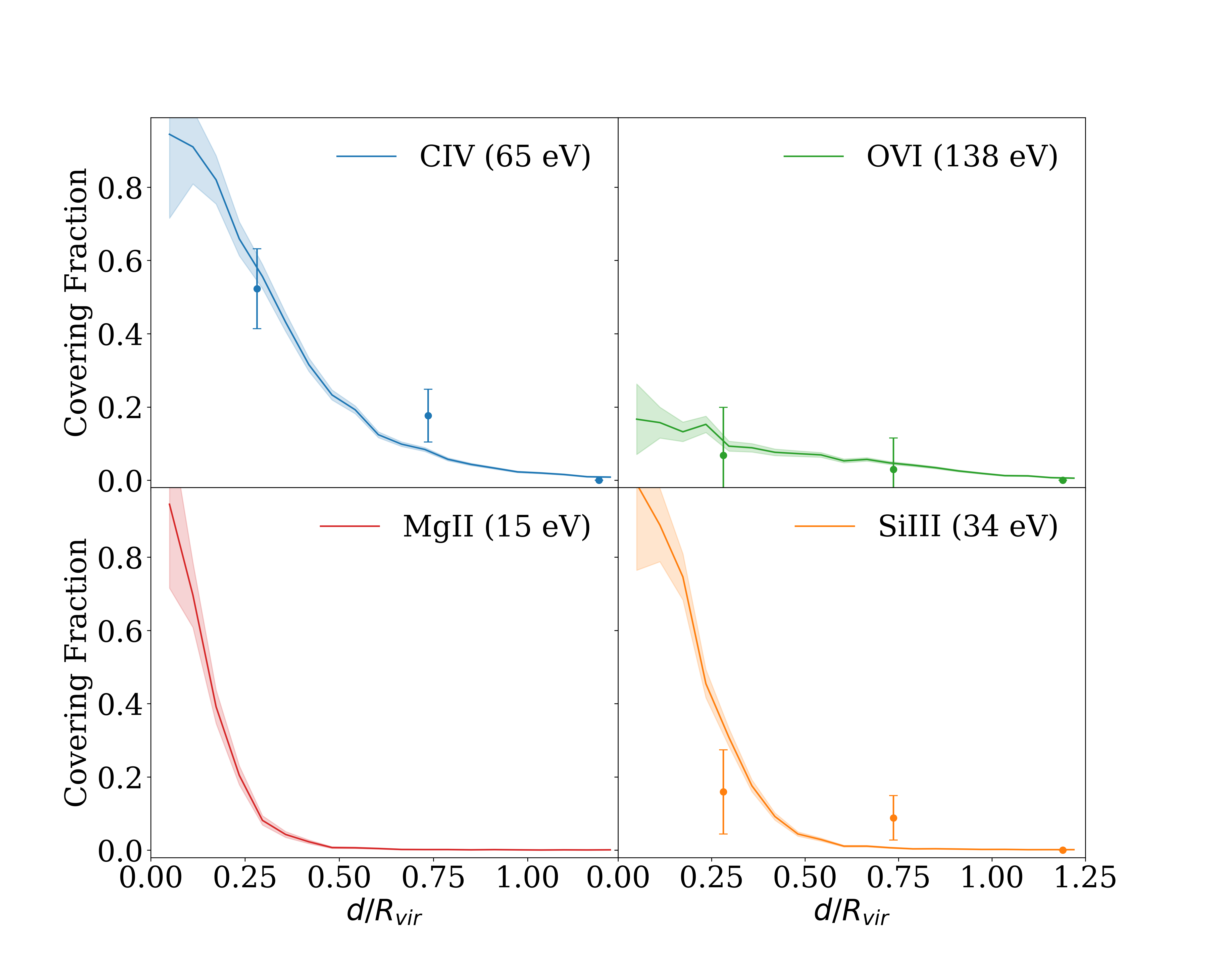}
    \caption{The covering fractions of \ion{C}{IV}, \ion{Si}{III} and \ion{Mg}{II} above 50 m\AA\,, and of \ion{O}{VI} above 80 m\AA\,, calculated from all selected LOS in the simulations, with the ionization energy for each state indicated. The different covering fraction threshold for \ion{O}{VI} is a result of the different detection limit for this ion in observations. The data points with error bars are compiled from the observations \citep{Bordoloi2014, Liang2014, Johnson2017}, which are binned radially. The covering fractions from the simulated sample are consistent with the results from observations.}
    \label{fig:cover_factor}
\end{figure*}

We examine the physical distribution of the ions and metals in the CGM based on the covering fractions and column densities of the ions.

The covering fraction of each given ion was calculated using the following formula:
\begin{equation}
    C_{f}(R)=\frac{N_{W}( \geq W_{cut}, R)}{N_{tot}(R)}
\end{equation}
where $N_{W}( \geq W_{cut}, R)$ is the number of LOS with absorbers with EW greater than the cutoff EW ($W_{cut}$) within an impact parameter $R$, while $N_{tot}(R)$ is the total number of LOS that we sample in the same region.

\autoref{fig:cover_factor} shows the covering fractions of \ion{C}{IV}, \ion{Si}{III} and \ion{Mg}{II} above 50 m\AA\,, and of \ion{O}{VI} above 80 m\AA\,, compared with the observations \citep{Bordoloi2014, Liang2014, Johnson2017}. The thresholds are chosen to match the lowest detection limits of the ions. The covering fractions from the simulated sample are in agreement with the results from observations.

\begin{figure*}
	\includegraphics[width=\textwidth]{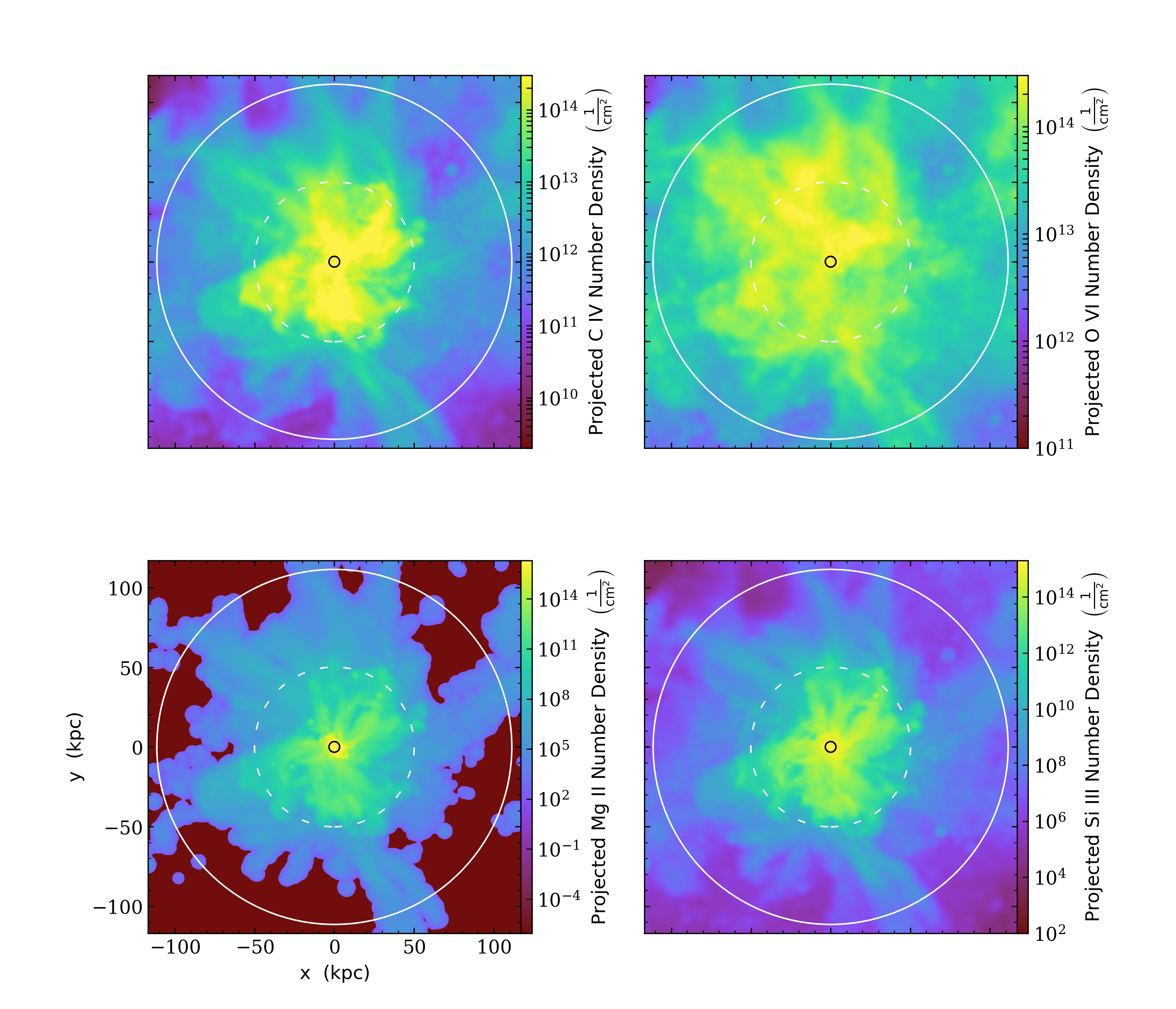}
    \caption{Column density maps of \ion{C}{iv}, \ion{O}{vi}, \ion{Mg}{ii}, and \ion{Si}{iii} at $z=0.061$ for simulation run `m11i'. The black circle shows the $80\%$ stellar mass radius, the dashed white circle shows the boundary between the inner and the outer parts of the CGM (0.45 $R_{\rm vir}$, the edge of the wind-dominated regime), and the solid white circle shows the virial radius. For \ion{C}{iv}, the column density map shows three distinct regions: the ISM with the highest \ion{C}{iv} column density, the inner part of the CGM with a lower \ion{C}{iv} column density compared with the ISM, and the outer CGM with the lowest \ion{C}{iv} column density; the column densities of \ion{O}{vi} in the three zones look comparable; for \ion{Mg}{ii}, there are also three distinct regions in the column density map: the ISM where the \ion{Mg}{ii} column density is the highest, the inner part of the CGM with lower \ion{Mg}{ii} column density, and the outer CGM where the \ion{Mg}{ii} column density is close to 0; the column density distribution of \ion{Si}{iii} is similar to that of \ion{Mg}{ii}. }
    \label{fig:CD_C}
\end{figure*}

We can also analyze the column densities of the ions directly in the simulations. The column density maps of the ions are shown in \autoref{fig:CD_C}. The three circles shown in the maps are the $80\%$ stellar mass radius, the boundary between the inner and the outer parts of the CGM (0.45 $R_{\rm vir}$ which we identify as the edge of the wind-dominated regime), and the virial radius. We arrived at this definition for the inner most radius as follows: we plotted the cumulative stellar mass inside the virial radius of the main halo. In all the halos, we saw a sharp break in the cumulative mass function at around $80\%$ of the total stellar mass in the halo. We call this the $80\%$ stellar mass radius, and we use this as the boundary between the ISM and the CGM.  

The column densities of the ions are different in the three regions: for \ion{C}{IV}, the ISM has the highest column density, the inner part of the CGM shows a lower column density compared with the ISM, and the outer CGM has the lowest column density; \ion{Mg}{II} and \ion{Si}{III} show similar trend as \ion{C}{IV}, except that the column densities of the low ions show more drastic decreases in the outer CGM; \ion{O}{VI} has similar column densities in all three regions. 

\begin{figure}
	\includegraphics[width=\columnwidth]{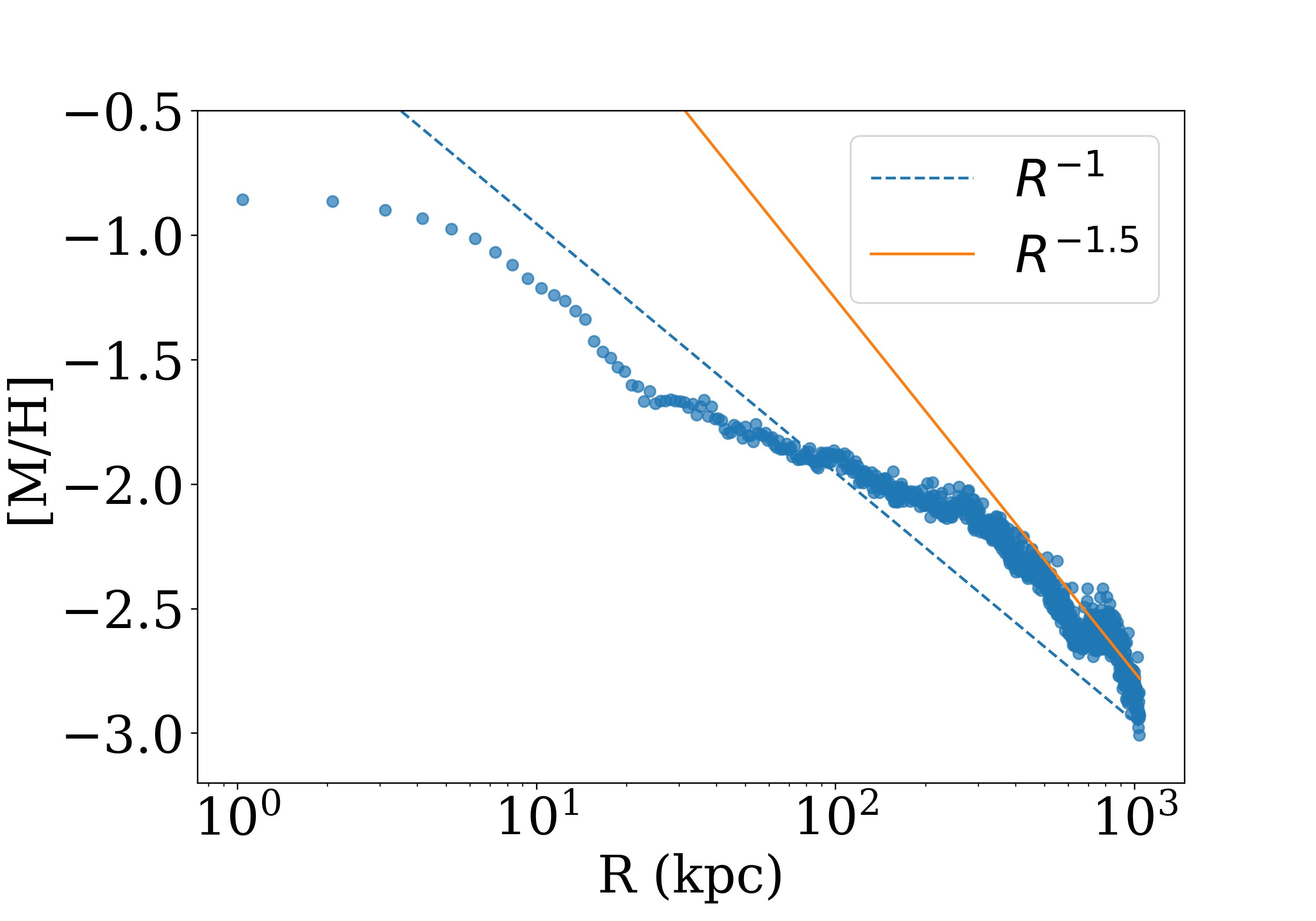}
    \caption{The run of metallicity with radius in simulaiton `m11b' at z=0.061. The blue and orange lines are power laws of index -1 and -1.5, as indicated in the legend.}
    \label{fig:metallicity}
\end{figure}

\begin{figure}
	\includegraphics[width=\columnwidth]{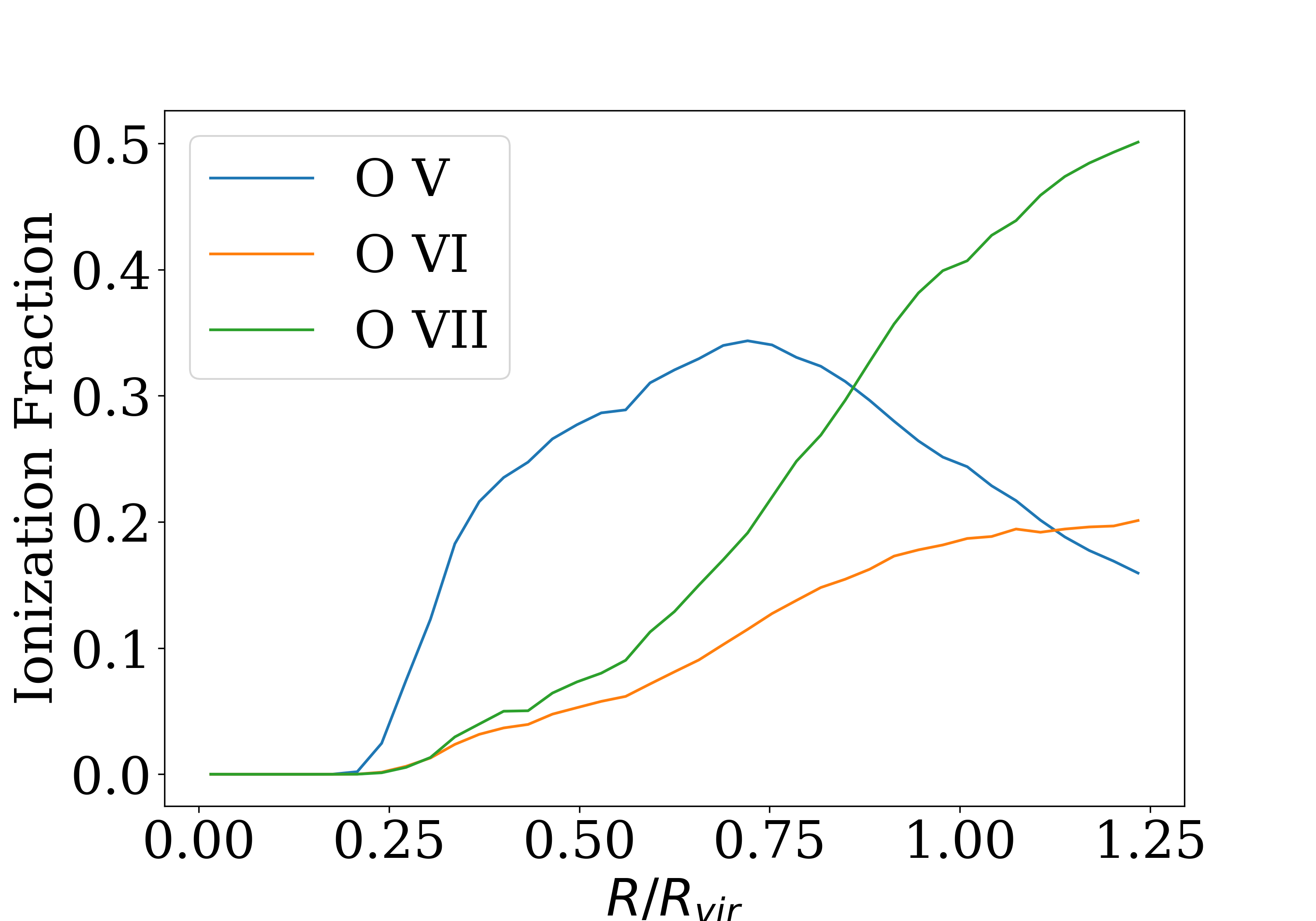}
    \caption{The ionization fractions of \ion{O}{v}, \ion{O}{vi}, and \ion{O}{vii} relative to oxygen in simulation run `m11b' at z=0.061. The profile of \ion{O}{vi} as a function of radius better traces the ionization state of oxygen, rather than its density.}
    \label{fig:ionization_fraction}
\end{figure}

The roughly constant column density of \ion{O}{VI} is consistent with the roughly constant cover factor that we see in \autoref{fig:cover_factor}. Why does \ion{O}{VI} behave so differently than the other ions that we study? \autoref{fig:metallicity} shows the run of metallicity with radius; it decreases roughly as a power law with radius. Since both the density (not shown) and metallicity drop off with increasing radius, one might expect the \ion{O}{VI} column density would drop off with radius as well. However, \autoref{fig:ionization_fraction} shows the three dominant ionization states of oxygen as a function of radius in a simulation snapshot (`m11b', z=0.061), demonstrating that the ionization fracgtion of \ion{O}{VI} increases with radius. Recalling that the column density is the product of the ion \ion{O}{VI} fraction times the density times the metallicity times the pathlength, the rapid increase in the \ion{O}{VI} ionization fraction is largely responsible for the relatively flat run of column density at large impact parameter in the \ion{O}{VI} panel in \autoref{fig:CD_C}.

As confirmation of this interpretation, we have examined the run of the $\ion{C}{IV}$ fraction (not shown) which, in contrast to the run of $\ion{O}{VI}$, decreases with increasing radius.

We see a boundary in the column densities of the ions at 0.45 $R_{\rm vir}$, which appears to separate an ``inner'' and ``outer'' region in the CGM. This is consistent with the location where the fraction of the total CGM mass contributed by wind and by IGM accretion is equal for $10^{11} \rm M_{\rm \odot}$ halos in FIRE-2 simulations \citep{Hafen2019}. 

\begin{figure*}
	\includegraphics[width=\textwidth]{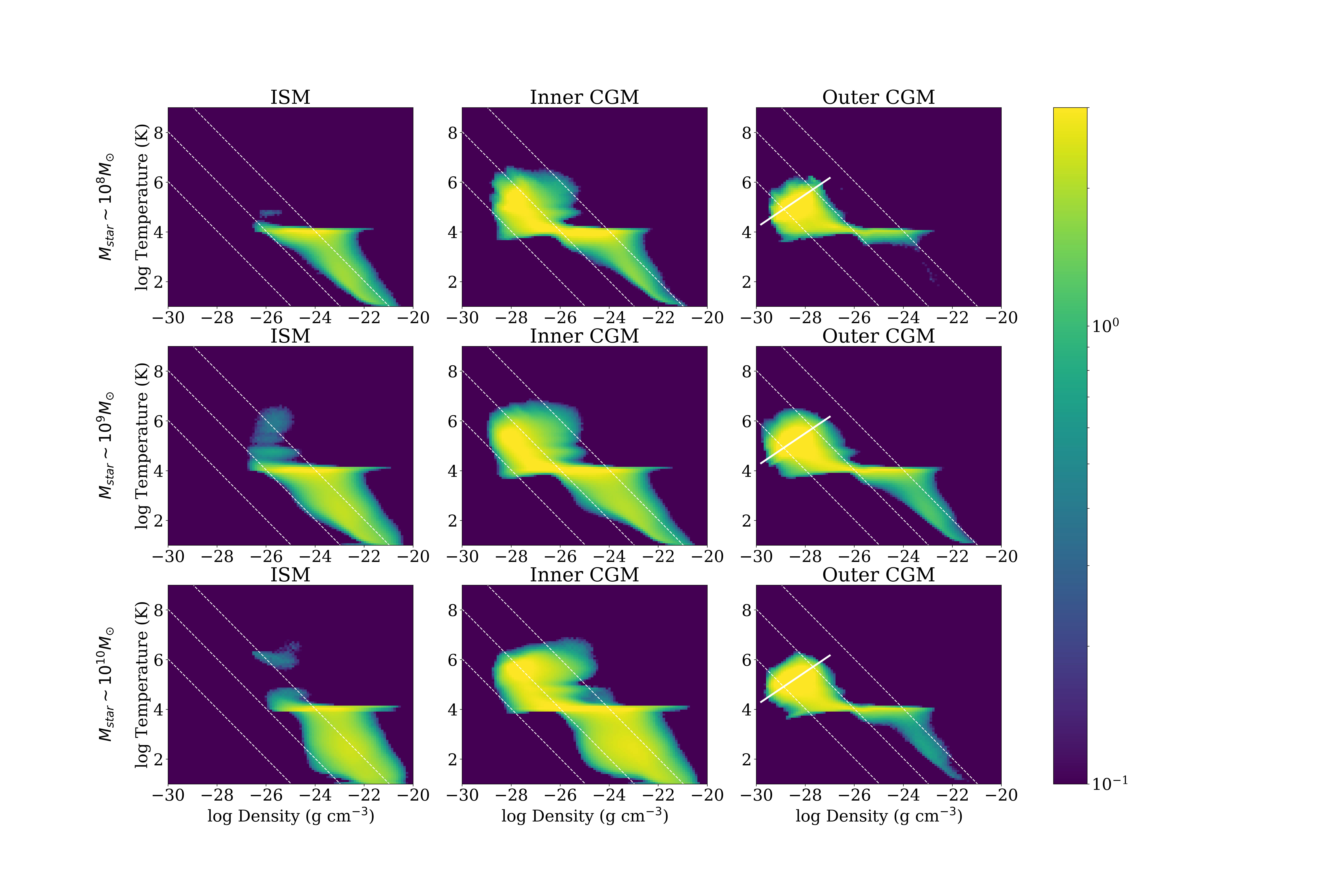}
    \caption{This figure shows the density -- temperature phase diagrams of three regions inside the virial radii of the simulated galaxies in three mass bins. The three rows show galaxies with stellar masses $10^{7.5}\,M_{\rm \odot}<M_{\rm *}<10^{8.5}\,M_{\rm \odot}$, $10^{8.5}\,M_{\rm \odot}<M_{\rm *}<10^{9.5}\,M_{\rm \odot}$ and $10^{9.5}\,M_{\rm \odot}<M_{\rm *}<10^{10.5}\,M_{\rm \odot}$. The three columns show the disky ISM ($R <0.03\,R_{\rm vir}$), inner CGM ($0.03\,R_{\rm vir}< R <0.45\,R_{\rm vir}$) and outer CGM ($0.45\,R_{\rm vir}<R<R_{\rm vir}$) regions surrounding the galaxies. Colors show the probability density weighted by gas mass. The gas in these three regions reside in different places in the density -- temperature diagram. The white dashed lines are isobaric lines with pressures of $1.0  \times 10^{-16}$,  $1.0 \times 10^{-14}$, and $1.0 \times 10^{-12}$ Ba. The white lines show the scaling $T \propto \rho^{2/3}$, which comes from the relationship when entropy is constant. The outer CGM follows the adiabats indicated by the white lines, while the inner CGM and the disky ISM clearly follow different trends. The fact that the high-temperature gas in the outer CGM (right-most column) traces an adiabat, shows that the gas in that region is well-mixed.}
    \label{fig:phase}
\end{figure*}

The density -- temperature diagrams for gas in the three regions (disky ISM, inner CGM and outer CGM) are shown in \autoref{fig:phase} in three stellar mass bins. The gas in these three regions resides in different places in the phase diagram: the ISM gas is cold and dense compared to gas in the inner and outer CGM. The outer CGM follows the adiabats ($T \propto \rho^{2/3}$) indicated by the white lines, while the inner CGM and ISM do not. This indicates different energetic states between the three regions, which may be the result of the different origins of the gas in these regions \citep{Hafen2019}.

\begin{figure*}
	\includegraphics[width=\textwidth]{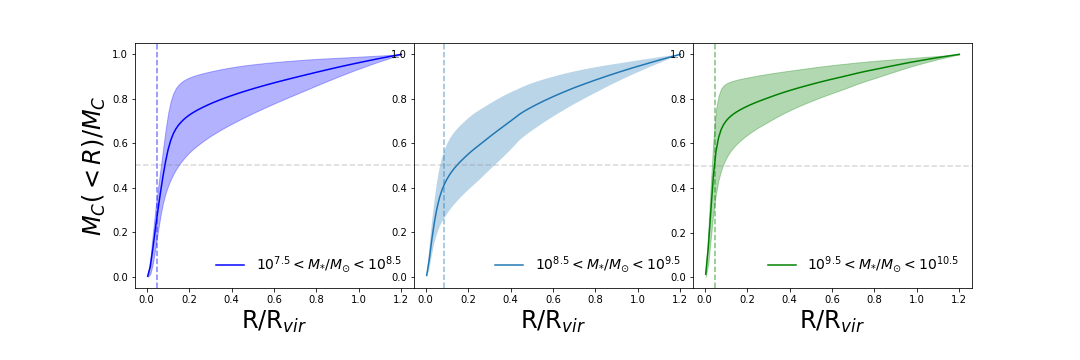}
    \caption{The cumulative carbon mass distribution of the simulated galaxy samples in three mass bins. The vertical dashed lines indicate the distance containing 80\% of the stellar mass for each bin, corresponding to $R/R_{\rm vir}$ values of 0.050, 0.084, and 0.047. We treat this as a dividing line between the ISM and CGM. For all mass bins, the carbon mass in the CGM is larger than or comparable to the carbon mass in the ISM. As the stellar mass of the galaxy decreases, a larger fraction of the carbon resides in the CGM.}
    \label{fig:carbon_mass_ISM_CGM}
\end{figure*}

We can compare the simulated carbon mass fraction to those inferred from observations. The cumulative carbon mass distribution of the simulated sample is shown in \autoref{fig:carbon_mass_ISM_CGM} in three mass bins. We use the 80\% stellar mass radius as the boundary between the ISM and CGM, as discussed earlier. For all three stellar mass bins, the carbon mass in the CGM is larger than or comparable to the carbon mass in the ISM, which is in agreement with the estimates in \citet{Bordoloi2014}. We also note that as the stellar mass of the galaxy decreases, a larger fraction of the carbon resides in the CGM of the simulated sample.

\begin{figure*}
	\includegraphics[width=\textwidth]{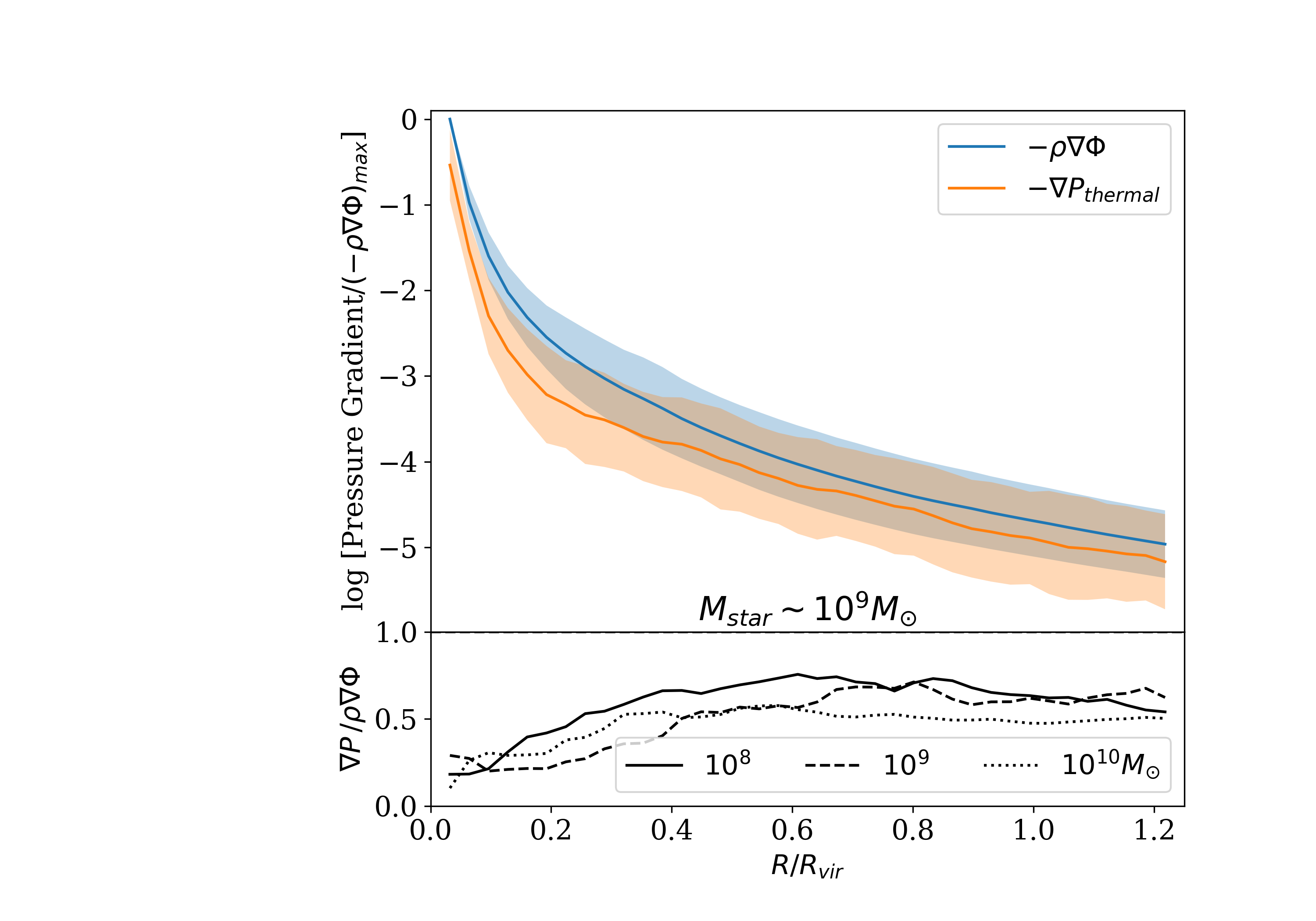}
    \caption{The upper panel shows the gas thermal pressure gradient $\nabla P_{thermal}$ radial profile (orange) compared to the gravitational force per unit volume $\rho \nabla \Phi$ (blue) profile. The solid lines are the mean of the logarithm values for simulation snapshots with central galaxies in the median stellar mass bin ($10^{8.5}\,M_{\rm \odot}<M_{\rm *}<10^{9.5}\,M_{\rm \odot}$), and the shaded regions are between the ${\sim}\pm$ 1 $\sigma$ values. The profiles are calculated by volume averaging the density and pressure over spherical shells centered at 40 values of $R/R_{\rm vir}$. The lower panel shows the ratio of thermal pressure to gravitational force as a function of normalized galacto-centric radius $R/R_{\rm vir}$ in three mass bins. The inner CGM is far from HSE. The outer CGM is nearly but not quite supported by thermal gas pressure, thus not in HSE.}
    \label{fig:he_09}
\end{figure*}

\subsection{Dynamics of the CGM}\label{sec:dynamics}

\begin{figure}
	\includegraphics[width=\columnwidth]{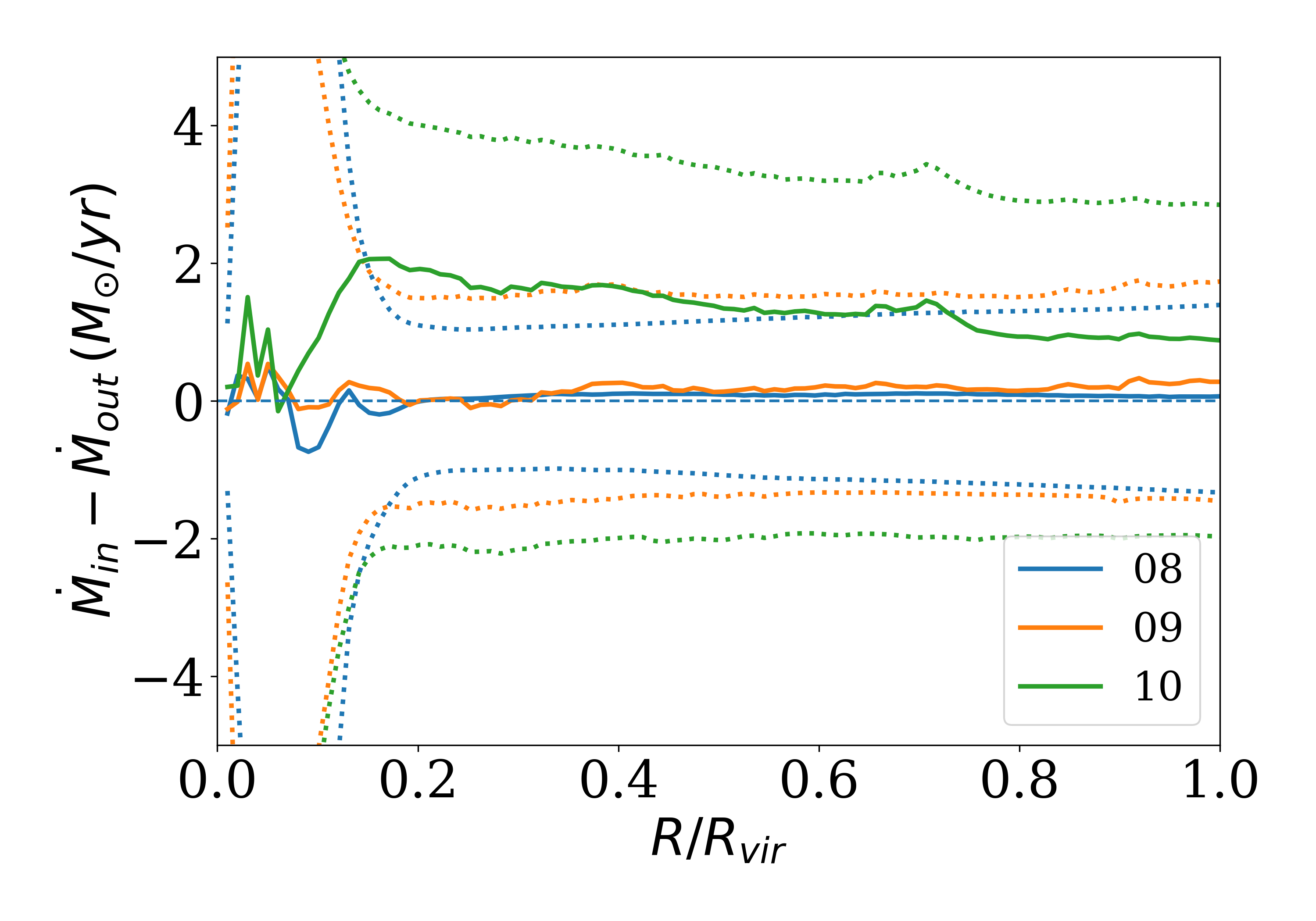}
    \caption{The average net inflow rates and gross inflow/outflow rates as a function of galacto-centric radius normalized by virial radii $R_{\rm vir}$. The dotted profiles above zero are gross inflow rates, the profiles below zero are gross outflow rates. The solid lines represent net mass transfer rates. The mass transfer rates are calculated for the simulated sample in three stellar mass bins: `08' (blue, $10^{7.5}\,M_{\rm \odot}<M_{\rm *}<10^{8.5}\,M_{\rm \odot}$), `09' (orange, $10^{8.5}\,M_{\rm \odot}<M_{\rm *}<10^{9.5}\,M_{\rm \odot}$), and `10' (green, $10^{9.5}\,M_{\rm \odot}<M_{\rm *}<10^{10.5}\,M_{\rm \odot}$). The net inflow rates are comparable to the SFR, and are an order of magnitude smaller than the gross inflow and outflow rates, which are comparable to each other. The net inflow rate in the `10' mass bin is dominated by simulation `m11f', which has a large SFR, if it is removed, the net inflow rate looks like the other two mass bins.}
    \label{fig:net_inflow}
\end{figure}

Throughout this paper, we have mostly focused on accounting for the mass budget of the multiphase CGM. However, the structure and composition of the CGM are directly related to its dynamics, which is critical to understanding how the CGM fuels the star-forming regions of dwarf galaxies. In this subsection, we examine how the composition of the CGM relates to its dynamics. Specifically, recent analytic models \citep[e.g.,][]{Mathews2017, McQuinn2018, Faerman2019} assume that the CGM of $L_{\rm *}$ galaxies is in hydrostatic equilibrium (HSE) or near HSE with a net inflow \citep{Stern2019}. Here we test whether the CGM of the simulated dwarf galaxies deviates from HSE by comparing gas pressure gradients to gravitational force. 

We show the gas thermal pressure gradient ($d P_{thermal}/dr$) and the gravitational force ($\rho (r) G M(r)/r^2$) as a function of galacto-centric radius in the top panel of \autoref{fig:he_09}. The profiles are calculated by volume averaging the density and pressure over spherical shells centered at 40 values of $R/R_{\rm vir}$. The thermal pressure gradient is calculated by dividing the pressure difference between adjacent shells by the distance between them. The bottom panel of \autoref{fig:he_09} shows the ratio of $d P_{thermal}/dr$ to $\rho (r) G M(r)/r^2$. The ratio ranges from 0.5 (for $M_{\rm star} \sim 10^{10} M_{\odot}$) to 0.7 (for $M_{\rm star} \sim 10^{8} - 10^{9} M_{\odot}$) for $R>0.4 R_{\rm vir}$, so the gas is nearly but not quite in HSE. At smaller radii, the ratio is smaller. This is consistent with the findings in \citet{Stern2020b}, who find that in halos with same mass as those we study, only gas at larger radii is in HSE supported by thermal gas pressure. In general, the outer CGM in our simulations is mostly supported by thermal pressure, and is near but not in HSE, which means that we expect to see significant motion in the medium (see \citet{Esmerian2020}, for similar results for Milky Way-mass galaxies simulated with the FIRE-2 physics). 

To investigate the CGM's deviation from HSE, we calculate the inflow and outflow rates for the three stellar mass bins as a function of galacto-centric radius, which are shown in \autoref{fig:net_inflow}. We then compare those rates to the net accretion rate (inflow - outflow). We see a net inflow through the CGM of order 0.2 $M_{\rm \odot}/yr$ for the median stellar mass bin. This inflow rate, which increases with stellar mass, is comparable to the SFR of the galaxy, also of order 0.2 $M_{\rm \odot}/yr$ for the median stellar mass bin. However, the gross inflow and outflow rates, while comparable to each other, are larger than the net inflow rate by an order of magnitude, i.e., $\dot M\approx 2 M_\odot\,{\rm yr}^{-1}$. 

We find that the mass-weighted velocities of the inflow and outflow (${\sim}$ 100 km/s) are similar to the virial velocity of the halo, also consistent with the ratio of the thermal and gravitational pressures we see. 
Beyond explaining the dynamics of the system, this picture has a number of consequences: first, it appears that the inflow and outflow rates are well coupled throughout the halo; secondly, this also implies that whatever mechanisms drive the outflows, they are effective throughout the galactic halo; and consequently, the dynamics of the halo are coupled to the star-formation activities in the central stellar region of galaxy.

\section{Discussion}\label{sec:discussion}
In this section we discuss the limitations in the simulations and our analysis, such as the resolution and physics in the simulations and the post-processing tool used, and how those affect the results presented in this paper. 

The typical gas mass resolution achieved for dwarf galaxies in FIRE simulations is ${\sim} 5 \times 10^{3} M_{\rm \odot}$ (\autoref{tab:parameters}). Depending on the clump size of the gas in the CGM, the current resolution may not be enough to resolve the dynamics in the dense CGM phases. There is observational evidence suggesting that some low-ion absorbers have sub-kiloparsec length scales \citep{Lehner2013, rudie2019column}. For a clump radius of 400 pc, \citet{Crighton2015} estimates the minimum required mass resolution is 4 $ \rm M_{\rm \odot}$, which is a substantially finer resolution  than available in cosmological zoom simulations like FIRE.
The low-ion content is likely to increase with improved mass resolution in the CGM, as suggested by recent work by other groups \citep{Peeples2019, Suresh2019, vandeVoort2019}.

We used the metagalactic radiation field \citet{Haardt2012} when calculating densities and ionization states of metals using Trident. However, we did not include local photoionizing sources like nearby stars in the galaxy. This affects the analysis in two ways: it is known that \citet{Haardt2012} under-predicts observed \ion{H}{I} photoionization coefficients by a factor ${\sim} 2$ \citep[see, e.g.,][]{Faucher2020}, so this likely has an effect on the prediction of the low ionization-state ions in this work. Including the stellar radiation will change the ionization states of the ions in the CGM. However, because the stellar radiation decreases radially as a function of 1/$r^{2}$, it is unlikely to change the distribution of ionization state much beyond the galactic disk.

The simulation suite described so far in this work does not include cosmic rays (CRs). \citet{Ji} studied the effects of CRs on galaxies in the FIRE simulations across a broad range of stellar and halo masses, and found that CR pressure appears to become important only when the galaxy halo mass approaches $10^{12}M_\odot$. CRs in the FIRE simulations do not appear to have a strong impact on the CGM of dwarf galaxies. 

In a first step at quantifying the effects of CRs on dwarf galaxies in FIRE,  we examined the CR version of the median stellar mass galaxy `m11e' in our sample.  We find no significant difference in the EW distributions of the ions between the two runs. The CR run has only ${\sim}44\%$ of the stellar mass compared to the non-CR run at low redshift, so we also compare the CR run `m11e' with a non-CR run `m11q' with comparable stellar mass at $z=0$, and again find no difference in the EW distributions in most ions, with the exception of \ion{O}{VI}, which has slightly higher EWs in the CR run.

If the CGM distributions around dwarf galaxies in the real universe resemble the results presented in this work below the detection limit, and given the detection thresholds of current instruments, observers probing the absorption properties of the low ions when the background QSO is beyond ${\sim} 0.4 R_{\rm vir}$ of the galaxy, and beyond ${\sim} R_{\rm vir}$ for \ion{C}{IV} will likely only observe upper limits given the current detection limits.

\subsection{Origins and Fates of Halo Metals}\label{sec:Origin}
Using the same suite of FIRE-2 simulations as in this work, recent works trace the origins and fates of the material composing the CGM \citep{Hafen2019, Hafen2019b}. These studies focus on halo masses $M_{h} (z = 0) {\sim} 10^{10}-10^{12} \rm M_{\rm \odot}$ and redshifts $z \,=\, 2$ and $z \,=\, 0.25$, and show that across the whole halo mass range and redshift range studied, the main origin ($\gtrsim 60 \%$) is IGM accretion (including infalling halos), with wind from the central galaxy being the second most important contribution of the CGM, while wind from satellite galaxies contribute the least. The exception is for $L_{\star}$ halos, where the contributions from winds of the central galaxy and satellite galaxies are about equal.

\citet{Hafen2019b} studies the fates of the CGM gas in the same halo and redshift ranges. For the CGM material at $z \,=\, 2$, immediately after leaving the CGM, most of the CGM gas enters the central galaxy; by $z \,=\, 0$, half of the mass remain in the ISM or stars in the central galaxy for the $M_{h} (z = 2) {\sim} 10^{11.5} \rm M_{\rm \odot}$ halos, but for lower mass halos most of the CGM mass are ejected into the IGM. For the CGM mass at z=0.25, most of the CGM mass remains in the CGM by $z \,=\, 0$ across the halo mass range studied.

In general, the two studies find that there is substantial mixing of the CGM gas, as a result, there is little correlation between the origins and the fates of the CGM material, and a typical LOS can intersect CGM gas of multiple origins.

\subsection{The Picture of CGM Composition and Dynamics Across Halo Mass and Redshift Ranges}

In this subsection we discuss the difference between this work and previous studies in stellar/halo mass ranges, redshift, dynamics, resolution, feedback prescriptions, and observables.

In this study we selected the stellar mass and redshift ranges to be representative of current observations of the CGM of dwarf galaxies. Not many previous studies focus on the CGM of simulated galaxies in the same mass and redshift ranges, and the limited overlap occurs at the high stellar and halo mass ends of this study (e.g., \citet{Ford2013}). \citet{Ford2013} study the CGM around $z=0.25$ galaxies in cosmological hydrodynamic simulations, and find the low ions (\ion{Mg}{II} and \ion{Si}{IV}) drop rapidly with impact parameter, while high ions (\ion{O}{VI} and \ion{Ne}{VIII}) have relatively flat radial profiles; and absorption increases with halo mass. These findings are in agreement with the trends of the ion absorption profiles presented in this work.

We concluded in Section \ref{sec:dynamics} that the CGM in this study is not in HSE. Most of the other studies find the same conclusion that CGM tends not to be in HSE in this halo mass range. \citet{Lochhaas2019} use spherically-symmetric, idealized, kpc-scale resolution simulations of the CGM, and found that the $10^{11} \rm M_{\rm \odot}$ halo's CGM is not in HSE, while the $10^{12} \rm M_{\rm \odot}$ halo contains hot gas in HSE with cold gas falling onto the galaxy. \citet{Oppenheimer2018b} tested whether the halos in the EAGLE zoom simulation are in HSE, and found that the Milky Way $L^{\star}$ halos deviates significantly from HSE, especially in the inner regions (see also \citet{Esmerian2020}, for FIRE-2 Milky Way-mass galaxies). In more massive halos ranging from $10^{11}$ to $10^{12} \rm M_{\rm \odot}$, \citet{Fielding2017} use idealized 3D hydrodynamics simulations to study the dynamics of the CGM, and find that below a critical halo mass $M_{crit}\, \approx\, 10^{11.5} \rm M_{\rm \odot} $, the CGM gas is not thermally supported, and above $M_{crit}$ the CGM is supported by thermal pressure created in the virial shock (see also \citet{Stern2020} for a generalization of this result for halos which are depleted of baryons).

The resolution in the diffuse CGM is usually orders of magnitude coarser than in the galaxy in the FIRE simulations due to the density-dependent nature of Lagrangian schemes.  Recent studies have increased the spatial resolution of the CGM in zoom simulations of Milky Way-like galaxies and have found substantial changes in the behavior and observables of the CGM relative to lower-resolutions. \citet{vandeVoort2019} used the AREPO code to progressively increase the spatial resolution of the CGM, finding order-of-magnitude increases in the column density of neutral hydrogen present in the CGM of the target galaxy. \citet{Hummels2019} and \citet{Peeples2019} both used the Enzo code to study a Milky-Way analog galaxy in zoom simulations with spatial resolutions up to an unprecedented 500 comoving parsecs throughout the CGM. Like the AREPO study, these groups also found an increase in \ion{H}{i} among other changes. \citet{Peeples2019} discovered significant differences in the kinematic structure of absorption line features in accompanying synthetic spectra for these galaxies. \citet{Hummels2019} explained the effect by identifying that increased resolution lends itself to preserving cool gas for longer in the simulation and and avoiding artificial mixing of hot and cold gas phases. In a related study, \citet{Schneider2018} present a set of 5-pc constant resolution simulations with isolated disk galaxies, and show that the simulations have not converged at 10 pc resolution, demonstrating the need for high resolution when studying the effects of winds on the CGM. Based on the results in these papers, improving the resolution in the halo of the simulations used in this work will likely increase the low-ion content, and would be critical if the goal is to resolve the internal kinematic structure of absorption.

Compared to the FIRE simulation suite, there are differences in the physics and feedback prescriptions used in other simulations studying the CGM. These differences can play a role in changing the predicted behavior and observables in the circumgalactic medium \citep{Hummels2013MNRAS.430.1548H}. For example, \citet{Lochhaas2019} use a simple formulation for galactic scale winds in the idealized simulations of the CGM; and \citet{Ford2013} use momentum-driven winds as the stellar feedback prescription in their cosmological simulations. The observables and tracers used in some studies differ from those used in this paper. For example, \citet{Augustin2019} use RAMSES zoom simulations to compute expected emission from the CGM, and find the predictions agree well with current observations after adjusting the \ion{Ly}{$\alpha$} escape fraction. The tracers can be studied using the FIRE simulation suite in future work.

\section{Conclusions}\label{sec:conclusions}
In this work, we use the FIRE simulation suite with physically motivated models of the multi-phase ISM, star formation and stellar feedback to study the CGM of dwarf galaxies at low redshift ($z \leq 0.3$). In direct comparison with observations, our main conclusions are as follows:

(i) The EW and cover fraction of \ion{C}{iv} predicted by the FIRE simulations are somewhat low compared to observational constraints inside of $0.5 R_{\rm vir}$ ($50 \pm 12 \%$ of the observations are above the detection limit, compared to only $27 \%$ of the simulated sightlines above the detection limit). At larger radii, the simulation results lie factors of several to tens or even hundreds below the observations. The equivalent widths of the other ions inside ${\sim} 0.5 R_{\rm vir}$ are also underpredicted, by factors between three to ten or higher. Beyond $0.5 R_{\rm vir}$, there are only non-detections from observations \citep{Bordoloi2014, Liang2014, Johnson2017} (\autoref{fig:EW}).

(ii) The FIRE simulations can explain the \ion{C}{iv} EW profile fits obtained from observations for dwarf galaxies in different $R_{\rm vir}$ bins, and can reproduce the EW profile fits when placing a similar detection limit on simulation data (\autoref{fig:civ_spatial}). 

(iii) The covering fractions from the simulated sample are in agreement with the results from observations (\autoref{fig:cover_factor}).

(iv) In the FIRE-2 simulations the majority of the carbon mass in gas in dwarf galaxies are in the CGM (\autoref{fig:carbon_mass_ISM_CGM}), which is in agreement with observations \citep{Bordoloi2014}.

As the simulated sample is in reasonable agreement with observations, we can make useful predictions with the simulation data to provide insight in the physics of the CGM around dwarf galaxies:

(v) All of the global properties ($M_{\rm *}$, $M_{\rm H}$, $M_{\rm *}$/$M_{\rm H}$, SFR, $R_{\rm vir}$, and sSFR) correlate with one another, except for sSFR; and CGM absorption lines correlate with all of the global property correlations, again, except for sSFR (\autoref{fig:corner_plot}). The correlation strength increases with the ionization potential of the ion. The EW of ions in our analysis do not show strong correlation with sSFR, impact parameter ($0<d<1.25 \, R_{\rm vir}$), or redshift ($0<z<0.3$) (see \autoref{tab:cc}).

(vi) We see a ``three-zone" structure in the CGM of dwarf galaxies, associated with the disky ISM, the inner CGM, and the outer CGM. The break between the inner and outer CGM is around 0.45 $R_{\rm vir}$, where the fraction of the total CGM mass contributed by wind and IGM accretion is found to be equal for $10^{11} M_{\rm \odot}$ halos in FIRE-2 simulations \citep{Hafen2019} (see \autoref{fig:CD_C}).
 
(vii) The flat radial profile of the \ion{O}{vi} column density with radius is driven by an increase in the fraction of oxygen in the \ion{O}{VI} state (\autoref{fig:ionization_fraction}). The column density of other ions show a decrease with radius (\autoref{fig:CD_C}), driven by the combination of decreasing density and metallicity with increasing radius.


(viii) The outer CGM in the simulations is nearly but not quite supported by thermal pressure, thus not in HSE, as demonstrated by the existence of halo-scale bulk inflow and outflow motions. The net inflow rates are comparable to the SFR; in the lower mass halos, the actual inflow and outflow rates are greater than the net inflow rate by an order of magnitude, and are comparable to each other and the virial velocity of the halo. This indicates that the feedback that drives the dynamics of the halo is coupled with the star-formation activity towards the centre of the galaxy (\autoref{fig:he_09} and \autoref{fig:net_inflow}).


\section*{Acknowledgements}

We thank Lauren Corlies for her help in using Trident. F.L. thanks Gunjan Lakhlani for sharing the routines to calculate the dynamical time. We benefited from discussions with Bili Dong and Maan H. Hani. The analysis on the FIRE simulation data were run on XSEDE computational resources (allocations TG-AST120025 and TG-AST120023) and on the CITA compute cluster Sunnyvale. This research was enabled in part by support provided by SciNet (http://www.scinet.utoronto.ca) and Compute Canada (www.computecanada.ca). Computations were performed on the Niagara supercomputer \citep[][]{Ponce2019arXiv190713600P, Loken2010JPhCS.256a2026L} at the SciNet HPC Consortium. SciNet is funded by: the Canada Foundation for Innovation; the Government of Ontario; Ontario Research Fund - Research Excellence; and the University of Toronto. The data used in this work were, in part, hosted on facilities supported by the Scientific Computing Core at the Flatiron Institute, a division of the Simons Foundation. This work was performed in part at the Aspen Center for Physics, which is supported by National Science Foundation grant PHY-1607611. CAFG was supported by NSF through grants AST-1517491, AST-1715216, and CAREER award AST-1652522, by NASA through grant 17-ATP17-0067, by STScI through grants HST-GO-14681.011, HST-GO-14268.022-A, and HST-AR-14293.001-A, and by a Cottrell Scholar Award from the Research Corporation for Science Advancement. 

Software: SCIPY \footnote{http://scipy.org}, NUMPY \footnote{http://numpy.org} \citep{Walt2011}, MATPLOTLIB \footnote{http://matplotlib.org} \citep{Hunter2007}, MPI4PY \footnote{http://pythonhosted.org/mpi4py/} \citep{Dalcin2005}, Trident \footnote{http://trident-project.org/} \citep{Hummels2017Trident}, and YT \footnote{http://yt-project.org} \citep{Turk2011}.

\section*{Data availability}

The data supporting the plots within this article are available on reasonable request to the corresponding author. A public version of the GIZMO code is available at http://www.tapir.caltech.edu/~phopkins/Site/GIZMO.html. Additional data including simulation snapshots, initial conditions, and derived data products are available at https://fire.northwestern.edu/data/.



\bibliographystyle{mnras}
\bibliography{CGM} 






\bsp	
\label{lastpage}
\end{document}